%% file: Meromorphic-final.tex
\documentclass[11pt,english,twoside]{article}
\pdfoutput=1
\usepackage{jheppub}
\usepackage{lscape}
\usepackage{amsmath}
\usepackage{slashed}
\usepackage{mathtools}
\input{mymacros}

\title{Meromorphic Flux Compactification}

\author[a]{Cesar Damian} 
\author[b]{and Oscar Loaiza-Brito}
\affiliation[b]{Departamento de Ingenier\'ia Mec\'anica, Universidad de Guanajuato, \\
Carretera Salamanca-Valle de Santiago Km 3.5+1.8 Comunidad de Palo Blanco, Salamanca, Mexico}
\affiliation[c]{Departamento de F\'isica, Universidad de Guanajuato, \\
Loma del Bosque No. 103 Col. Lomas del Campestre C.P 37150 León, Guanajuato, Mexico.}
\emailAdd{cesaredas@fisica.ugto.mx}
\emailAdd{oloaiza@fisica.ugto.mx}

\abstract{
We present exact solutions of four-dimensional Einstein's equations related to Minkoswki vacuum  constructed from Type IIB string theory with non-trivial fluxes. Following  \cite{Candelas:2014jma,Candelas:2014kma} we study a non-trivial flux compactification on a fibered product  by a four-dimensional torus and a two-dimensional sphere punctured by 5- and 7-branes. By considering only 3-form fluxes and the dilaton, as functions on the internal sphere coordinates, we show that these solutions correspond to a family of supersymmetric solutions constructed by the use of G-theory. Meromorphicity on functions constructed in terms of fluxes and warping factors guarantees that flux and 5-brane contributions to the scalar curvature vanish while fulfilling stringent constraints as tadpole cancelation and Bianchi identities. Different Einstein's solutions are shown to be  related by U-dualities.  We present three supersymmetric non-trivial Minkowski vacuum solutions and compute the corresponding soft terms. We also construct a non-supersymmetric solution and study its stability. 
}

\arxivnumber{}


\begin{document}

\maketitle

\flushbottom 
\newpage

\section{Introduction}
String flux compactification has been extensively studied in the last decade opening up a strength relation among geometry and the construction of stable four-dimensional vacuum solutions. We now understand that they are geometrical extensions of those exact solutions constructed in the absence of fluxes corresponding to Calabi-Yau compactifications. Even more, flux compactification solves the so called moduli stabilization problem and gives us stringent insights about the possibility to construct four-dimensional de Sitter (dS) vacua. At the present stage it is  commonly accepted that dS vacua can be gathered from a flux compactification in the presence of orientifold planes and anti D3 branes\footnote{See for instance \cite{CaboBizet:2016qsa, Andriot:2016xvq} for  recent discussions on possible classical extra constraints.}\cite{Giddings:2001yu, Kachru:2003aw} or by the inclusion of non-geometric fluxes \cite{Damian:2013dq, Damian:2013dwa, Blumenhagen:2015xpa}.  Also, it is well known that for supersymmetric flux backgrounds, Einstein's equation is satisfied if we demand Bianchi identity and supersymmetry. Although flux compactification brings a consistent scenario for dS (see for instance \cite{Dasgupta:2014pma, Junghans:2016uvg} for some interesting discussions), the presence of localized sources introduce singular points at which the fluxes, for most of the cases \cite{Giddings:2001yu},  have not an analytical expression and are not exact solutions of the equations of motion. This is a consequence of taking trivial fluxes  (not depending on internal coordinates or moduli), an assertion valid only in a dilute flux limit.\\ 

Another problem faced by flux string compactification involves reproducing a minimal supersymmetric effective theory in four-dimensions which could embed a supersymmetric extension of the Standard Model of particles while preserving chirality and solving the hierarchy problem for the Higgs boson. However, as result of the last experiments run in the LHC, the possible presence of supersymmetry at low scales as TEV's is close to be discarded and therefore, supersymmetry appears to be non essential for solving the hierarchy problem. Although such a problem remains unsolved it opens up the possibility to consider non-supersymmetric string compactifications, i.e., models on which susy breaking scale is close to the string compactification scale \cite{Blumenhagen:2015kja}:
\begin{equation}
M_p> M_s>M_{KK}> M_{\text{comp}}\sim m_{\text{SUSY}}> m_{\text{inflaton}}.
\end{equation}
Hence, it is desirable to consider more generic flux scenarios which allow us to face these kind of problems. One possibility concerns turning on non-constant fluxes.
Compactification in the presence of fluxes depending on the internal coordinates or moduli have been considered previously \cite{Greene:1989ya,Dasgupta:1999ss}, while examples of U-folds with flux in string theory and M-theory were studied in \cite{McOrist:2010jw}. Further studies on non-trivial flux compactifications were considered in \cite{Martucci:2012jk,Braun:2013yla,Lust:2015yia}. More recently, it was constructed a family of exact solutions of compactifications threaded by fluxes depending on internal coordinates by \cite{Candelas:2014jma,Candelas:2014kma} and sourced by branes of diverse dimensionalities. Specifically the authors show that for a compactification on a fibered internal space given by a warped product of a four-dimensional torus  and a punctured sphere, it is possible to satisfy Bianchi identity and supersymmetric conditions for suitable choices of meromorphic functions depending on the complex coordinates of the sphere. In the same way as F-theory, these flux compactifications with meromorphic functions, dubbed G-theory,  gives a geometrical interpretation of the U-duality group (see \cite{Kumar:1996zx,Larfors:2015umb}), by replacing the tori by an auxiliary K3.\\

In this work we study generic conditions upon which a flux configuration depending on the same internal coordinates of the sphere, with a similar compactification on $T^4\times _z \mathbf{S}^2$, satisfy Einstein's equations. By turning on  3-form fluxes and the dilaton, sourced by 5- and 7-branes respectively, we find that a family of solutions of Einstein's equations are given precisely by meromorphic functions on $\mathbf{S}^2$.  For that, we have make an extensive use of the Global Residual Theorem \cite{Griffiths:1979pa} in complex analysis, which states that on a compact space with singular points the total sum of residues related to meromorphic functions vanishes. This allows us to prove that by the simple use of ``meromorphic fluxes" on the sphere, $-$meaning that we construct meromorphic functions formed by non-trivial closed string potentials and warping factors$-$, it is possible to satisfy Bianchi identities, tadpole conditions and Einstein's equations,  circumventing some results followed by the well known no-go theorem \cite{Giddings:2001yu} as having a constant warping factor in the absence of a five-form flux or the necessity to have orientifold 3-planes to obtain a Minkowski vacuum  with no one-form fluxes. Moreover, these family of solutions are compatible with those constructed by the same method described in \cite{Candelas:2014jma,Candelas:2014kma} by the used of G-theory. We also find that different solutions to Einstein's equations are related by U-dualities.\\

Once we show this, we focus on the construction of supersymmetric solutions by the use of U-dualities in the same context as the family of solutions studied in \cite{Candelas:2014jma,Candelas:2014kma}. Based on these solutions, we construct a non-supersymmetric solution in the effective space-time which is stable for some  regions on the complex coordinates of the internal sphere, at least at leading order with respect to the location of singularities. Although we consider this an important step through the construction of dS vacua with possible implications on cosmological issues, we leave this important task for future work.\\

Our work is organized as follows: In section 2 we review the flux configuration taken through all our study as well the construction of the punctured two-dimensional sphere by following the results studied in \cite{Candelas:2014jma,Candelas:2014kma}. After that we concentrate on solving  Einstein's equations. As mentioned, we make an extensive use of the Global Residue Theorem. Since our flux configuration break half of supersymmetries in the effective theory, we also review the calculation of the corresponding soft terms from the DBI action of a set of D5 branes warping a 2-cycle of the compact space \cite{Camara:2003ku,Camara:2004jj,Camara:2014tba} which precisely source the 3-form RR fluxes. In section 3 we show three supersymmetric solutions, compatible with Bianchi identities, fulfilling the supersymmetric conditions and compatible with the Einstein's solutions presented in the previous section. Finally, in section 4 we construct a non-supersymmetric solution based on the supersymmetric solutions and show their linear stability assuming spherical symmetry in the four-dimensional space-time. In section 5 we present our conclusions about the use of  ``meromorphic fluxes".  As reference, in appendix A we show our index notation and for completeness we review some useful gamma algebra in Appendix B, while in Appendix C  the non-zero components of the spin connection, necessary to solve the supersymmetric conditions, are given. Appendix D is devoted to present and prove the Global Residue Theorem. In Appendix E we show the dimensional reduction  of the Dirac-Born-Infeld action of $D5$-branes into the DBI action of induced $D3$-branes, required to compute the corresponding soft terms.

%---------------------------------------------------------------------------------------------------------------------------------------------------------------------
%---------------------------------------------------------------------------------------------------------------------------------------------------------------------
\section{Flux compactification}
For simplicity, we are interested in studying consistent string compactifications in Type IIB threaded with non-constant 3-form fluxes plus a non-trivial dilaton field. In such scenario, the  low energy action  for the bosonic sector in the string frame reads (see Appendix A for notation)
\be \label{eq:typeIIB}
S = \frac{1}{2 \kappa^2_{10}} \int d^{10}x \sqrt{-g}\left( e^{-2\phi} \left[ R + 4 ( \del \phi )^2 - \frac{1}{12} H_3^2 \right] -\frac{1}{12} F_{3}^2 \right)
-\frac{1}{4 \kappa_{10}} \int C_4 \w H_3 \w F_3 \, ,
\ee
where
\be \label{eq:mod-rr}
H_3 = {\rm d}B_2 \quad F_{3} = {\rm d} C_{2}  \, .
\ee
This action is symmetric under $SL(2,\mathbb{R})$ keeping the metric invariant while the doublet  $(C_2, B_2)$ and the complex dilaton $S = e^{-\phi} - {\rm i} C_0$ transforms  as
\be \label{eq:Sdual}
S \rightarrow \frac{a S - {\rm i} b}{{\rm i}c S + d} \, , \quad 
\left(
\begin{array}{l}
C_2\\
B_2
\end{array}
\right)
\rightarrow
\left(
\begin{array}{ll}
a & b\\
c & d
\end{array}
\right) \left( \begin{array}{l}
C_2\\
B_2
\end{array}\right) \, ,
\ee
for $ad-bc =1$. For $a = d = 0$ and $b = -c = 1$ the above symmetry reduces to S-duality under which the complex 3-form
\be \label{eq:G3def}
G_3 = F_3 - {\rm i} S H_3 \, ,
\ee
transforms accordingly. The  Killing Spinor Equations (KSE) read  \cite{Grana:2005jc,Blumenhagen:2013fgp}
\bea \label{eq:Killing_2}
\delta \Psi_M =  
\nabla_M \epsilon- \frac{1}{4} \slashed{H}_{M}  \sigma^3 \epsilon+\frac{e^\phi}{16}\slashed{F}_3 \Gamma_M \sigma^1 \epsilon
\eea
\bea \label{eq:Killing_1}
\delta \lambda = 
\left( \slashed{\del} \phi - \frac{1}{2} \slashed{H} \sigma^3 \right) \epsilon+\frac{e^{\phi}}{8}\slashed{F}_3 \sigma^1\epsilon \,,
\eea
where 
$\epsilon$ contains the two Majorana-Weyl
spinors of the same chirality, and $\sigma^1$ and $\sigma^3$ are the corresponding Pauli matrices.
The fluxes
are contracted as usual with the Gamma matrices $\Gamma^M$ in the curved background as
\be
\slashed{H}_M = \frac{1}{2} H_{MNO}\Gamma^{NO} ,\quad \slashed{\mathcal{G}} = \frac{1}{3 !} \mathcal{G}_{MNO}\Gamma^{MNO} \, ,
\ee
where $\mathcal{G}$ stands either for $F_3$ or $H_3$.
By taking a compactification on a six-dimensional space,
 $\Gamma^M$ decomposes into internal and external components as
\be
\Gamma^{\mu} = \gamma^{\mu} \otimes \mathbf{1} ,\quad \Gamma^{m} = \gamma^5 \otimes \gamma^{m} \,,
\ee
with the ten dimensional spinors $\epsilon^{1,2}$ decomposing as
\be
\epsilon^a = \xi_+ \otimes \eta_+^a + \xi_- \otimes \eta_-^a \, ,\nn \\
\ee
where $\xi$ and $\eta^a$ , $a=1,2$, stands for the components of the chiral spinors in the internal 
and compact space respectively, with the subscripts $\pm$ representing the corresponding chirality.  
It is useful at this point to define the  2-tuples $\eta$ and $\tilde\eta$ as
\be
\eta = \{\eta_+^1 , \eta_+^2  \} \quad \tilde \eta = \{\eta_+^2 , -\eta_+^1 \}\, ,
\ee
which we will use to denote solutions to the KSE.
%-----------------------------------------
\subsection{The anzatz}

In this work we are interesting in constructing type IIB supersymmetric (SUSY) and non-supersymmetric (non-SUSY) solutions by compactifying the internal six dimensions on a $T^4$ fibration over $\mathbf{S}^2$ in the presence of non-trivial 3-form fluxes.   For that, let us start by considering a non-compact  six-dimensional space consisting on the warped product $T^4 \times_z \mathbb{C}$. The 10-dimensional metric is given by 
\be \label{eq:manzats}
ds^2 = e^{2 A(z,\bar{z})} \eta_{\mu \nu}dx^{\mu}dx^{\nu}+g_{ij}(z,\bar{z}) dy^j dy^j + \Omega^2(z, \bar{z})dz d\bar{z},
\ee
where the fluxes, as well as the torus metric, dilaton $\phi$  and warping factors are taken to vary only over the complex plane and $i,j:1,\cdots , 4$ run over $T^4$ coordinates. In this background,  the internal Killing spinor equation for the gravitino, Eq. (\ref{eq:Killing_2}) decomposes into two equations related to $\mathbb{C}$ and  $T^4$  as
\be \label{eq:KSE-int1}
 \slashed{\del}A \eta^a - \frac{1}{4}e^{\phi} \slashed{F} \tilde \eta^a = 0 \, ,
\ee
\be \label{eq:KSE-int2}
\left( \nabla_{i} + \frac{(-1)^a}{4}H_i \right) \eta^a - \frac{(-1)^a}{8}e^{\phi}\slashed{F} \gamma_i \tilde \eta^a = 0 \, ,
\ee
while the corresponding dilatino variation, Eq. (\ref{eq:Killing_1}) reads
\be \label{eq:KSE-int3}
\left( \slashed{\del} \phi + \frac{(-1)^a}{2} \slashed{H} \right) \eta^a - \frac{e^{\phi}}{2}\slashed{F} \tilde \eta^a = 0 \, ,
\ee
for $a = 1,2$.  The RR and NS-NS 3-form fluxes $F_3$ and $H_3$ are the field strengths related to the 2-form potentials
\begin{eqnarray}
B_2&=&\sum_{i,j}b_{ij}(z,\bar{z})dy^i\wedge dy^j,\nonumber\\
C_2&=&\sum_{i,j}c_{ij}(z,\bar{z})dy^i\wedge dy^j,\label{eq:ByC}
\end{eqnarray}
with $b_{ij}$ and $c_{ij}$ real functions on $z$ and $\bar{z}$. \\

Our next step is to consider a  compactification threaded by these potentials. To do that, it is necessary to curl up the complex plane into a compact two-dimensional space. As studied in \cite{Greene:1989ya} this is accomplished by adding branes. Under their presence, the internal geometry backreacts  generating a deficit angle of $\pi /6$  for each brane, where for the critical case of 24  branes the complex plane curls up  into a two-dimensional punctured sphere.  In the background described by the metric (\ref{eq:manzats}) we can consider the presence of 3-, 5- and 7-branes wrapping 0-, 2- or 4-cycles in $T^4$ respectively while extending on the whole four-dimensional space-time. These branes are expected to induce  shift symmetries on the corresponding potentials by encircling the singular points in $\mathbf{S}^2$. Since we are considering only the presence of 3-form fluxes and the dilaton $\phi$, only five and seven branes are going to be taken into account.\\

Under this approach, the Bianchi identities have to be supplemented with delta-like local sources to compensate the flux. Therefore, our anzatz involves the presence of branes located at singular points $z_p$ on $\mathbb{C}$ making it  to curl up  to $\mathbf{S}^2$ such that the dilaton $\phi$ and the RR and NS-NS potentials $C_2$ and $B_2$  fulfill Einstein equations, Bianchi identities and the Killing equations of motion (for a SUSY case) while the total 5-brane must vanish on the internal space. \\

As said, the presence of a bound state of $q_1$ D5-branes and $q_2$ NS5-branes
induces a shift symmetry of the type $b_{ij} \rightarrow b_{ij} +q_1$ and $c_{ij} \rightarrow c_{ij} + q_2$ for $q_1 \, , q_2 \in \mathbb Z$, implying that, at leading order on $z_p$,  RR and NS-NS potentials behave as $\log \left(z-z_p \right)$ with 5-branes located at the singular points $z_p$ in $\mathbf{S}^2$.
Finally, as observed in \cite{Greene:1989ya}, U-duality invariance  completely fixes $\Omega(z,\bar{z})$ as 
\be
\Omega(z,\bar{z})=2e^{2D(z,\bar{z})}|h(z,\bar{z})|^2,
\ee
with
\be
h(z) =  \frac{\eta(U_1)^2 \eta(U_2)^2}{\Pi_{p=1}^{12} (z-z_p)^{1/12} \Pi_{p=1}^{12} (\bar z- \bar z_p)^{1/12} } \,,
\label{h}
\ee
where $\eta$ is the Dedenkin eta function with zeros at the points $z=z_p$ at which the branes are located; the terms in the denominator are added to remove those zeros  \cite{Greene:1989ya} such that the metric (\ref{eq:manzats}) in $\mathbf{S}^2$ is regular at every point \footnote{It is possible to merge the contribution of $D(z,\bar{z})$ and $|h(z,\bar{z})|^2$ into a single real function, however in compactifications with varying dilaton along $\mathbf{S}^2$, such real function is always factorized as a product of a modular invariant function and a real function \cite{Grana:2001xn}.}. The functions $U_{1,2}(z)$ are fixed to be meromorphic functions on the punctured sphere according to \cite{Greene:1989ya}. As we shall see, this  is accomplished by constructing complex functions in terms of real functions as the dilaton, fluxes and warping factors. In the next section we shall conclude similarly by solving Einstein equations for the internal compact space.

%--------------------------------------------------------------------------------------------------------------------------------------------

\subsection{Einstein equations}
Now we want to study $-$in the SUGRA context$-$ constraints on the previous flux configuration leading to  a Minkowski vacuum. 
Since in our setup, the internal fluxes are not constant but functions of internal coordinates, we want to show that by selecting flux potentials as appropriate real parts of holomorphic functions, we can construct a consistent compactification scenario where all geometrical and dynamical constraints are fulfilled while keeping internal fluxes as functions on complex coordinates on the punctured sphere as shown in \cite{Candelas:2014jma, Candelas:2014kma} where KSE and Bianchi Identity constraints were satisfied. \\

Let us start by considering the ten-dimensional Einstein equation in the string framework deduced from Type IIB SUGRA action,
\be
e^{-2\phi}\left(R_{MN}-\frac{1}{2}G_{MN}R\right)=T_{MN},
\ee
where $T_{MN}$ has  contributions from fluxes and local sources denoted as $T^f_{MN}$ and $T^l_{MN}$ respectively. In the trace reversed form, the 4-dimensional components of Einstein equations read\footnote{Throughout this work we use
\be
\nabla^2 A = \frac{1}{\sqrt{g}} \del_M \left( \sqrt{g} g^{MN} \del_N A \right) \,. \nn
\ee}

\be
e^{-2\phi}R_{\mu\nu}= T_{\mu\nu}-\frac{1}{8}T^L_Le^{2A}\eta_{\mu\nu}=e^{-2\phi}\left(R_{\mu\nu}(\eta)-\frac{1}{2}\eta_{\mu\nu}e^{-2A}\nabla^2 e^{2A}\right),
\ee
related to the metric (\ref{eq:manzats}). In a conformal Minkowski vacuum, by contracting with $\eta_{\mu\nu}$ one gets that
\be
e^{-2\phi}\nabla^2 e^{2A}=-\frac{1}{2}\hat{T} e^{2A},
\ee
where $2\hat{T}= T_\mu^\mu-e^{2A}T_m^m$. By considering constant fluxes it is well known  that $\hat{T}$ must vanish after integrating the above expression on a compact space \cite{Giddings:2001yu}. Since 3-form fluxes and D5-branes contribute positively to $\hat{T}$ it is concluded that objects with negative tension  must be considered in order to have a Minkoswki vacuum. In the present proposal, we are considering fluxes and warping factors depending on complex coordinates on $\mathbf{S}^2$ for which the above statements do not follow straightforwardly.\\

In the presence of only 3-form fluxes, the internal trace $T_m^m$ vanishes and 
\be
\hat T^f= -\frac{1}{3}e^{2A}\left(F_3^2+e^{-2\phi}H_3^2\right),
\ee
and therefore,
\be
e^{-2\phi}\nabla^2 e^{2A}=\frac{1}{6}e^{4A}\left(F_3^2 + e^{-2\phi}H_3^2\right)-\frac{1}{2}e^{2A}\hat{T}^l,
\ee
where $\hat{T}^ {f}$ is the contribution of internal fluxes while $\hat{T}^{l}$ is the contribution of 5-branes located at the singularities in the two-dimensional sphere.
Integration of $e^{-2\phi}\nabla^2 e^{2A}$ on the internal space $\mathbf{S}^2$  vanishes for an appropriate selection of non-constant warping factors and dilaton in the metric (\ref{eq:manzats}). To prove it,  we are going to make an extensive use of the {\it Global Residue Theorem} (GRT) \cite{Griffiths:1979pa} which establishes that summation of all residues of a holomorphic function vanishes on a compact space with singularities (see Appendix \ref{GRT} for details). A pedestrian way to understand this theorem is to consider that on a compact space as $\mathbf{S}^2$ a closed curve encircling all singularities $-$at which 5-branes are located$-$ can be deformed into a curve enclosing no singularities at all. Therefore, due to Stoke's law, residues of any holomorphic function on a compact space with singularities sum up to zero. \\

Hence, by integrating on the internal space the contribution to the trace of the energy-momentum tensor, we have that
\be
\int\sqrt{g_6}e^{-2\phi}\nabla^2 e^{2A} d^4ydzd\bar{z}=\int_{T^4\times \mathbf{S}^2} e^{-2\phi}d\ast_6 d e^{2A}
\label{eq:nablaA}
\ee
with 
\be
\ast_6 de^{2A}= \sqrt{g_6}\left(\partial_ze^{2A}dz + c.c\right)\wedge d({\cal V}_4),
\ee
and $g_6$ the determinant of the metric of $T^4\times _z \mathbf{S}^2$. Notice that the differential volume $d({\cal V}_4)= d^4y$ does not depend on the complex coordinates $z$ and $\bar{z}$. Now, by defining the real 1-form $\zeta_1= \partial_ze^{2A}dz $ we have
\be
\int_{T^4\times \mathbf{S}^2} e^{-2\phi}d\ast_6 d e^{2A}={\cal V}_4\int_{\mathbf{S}^2}\sqrt{g_4}e^{-2\phi} d(\zeta_1+ \bar\zeta_1),
\ee
where $g_4$ is the determinant of the 4-dimensional torus metric. Let us now assume there are $s$ singularities on $\mathbf{S}^2$ such that the space $S_\ast^2$ is defined as
\be
S_\ast^2=\mathbf{S}^2-\sum_{p=1}^s{\cal U}_p(z_p),
\ee
where $z_p$ is the location of the singularities on the complex sphere and ${\cal U}_p$ a region around $z_p$. It follows then that
\be
\int_{T^4\times \mathbf{S}^2} \sqrt{g_4}e^{-2\phi}d\ast_6 d e^{2A}={\cal V}_4\sum_{p=1}^\lambda\oint_{\partial{\cal U}_p} \sqrt{g_4}e^{-2\phi}(\zeta_1 +\bar\zeta_1)={\cal V}_4\sum_{p} Res(\sqrt{g_4}e^{-2\phi}\zeta_1(z_p)+c.c),
\ee
which vanishes for a closed $\zeta_1$ on $S_\ast^2$ $-$or equivalently if $\partial_zA$ is holomorphic$-$ and for $\nabla^2(\sqrt{g_4}e^{-2\phi})=0$ which is satisfied by $\partial_z g_4$ and $\partial_z\phi$ holomorphic on $S^2_\ast$.  Notice that this happens even for a non-trivial warping factor $A$. Settling down these functions we see that a consistent flux compactification to a Minkowski vacuum must satisfy 
\be
\int \sqrt{g_6}~ e^{4A}\left(\hat{T}^{f}+\hat{T}^{l}\right) d^4ydzd\bar{z}=0.
\label{Tf+Tl}
\ee
 Notice that for constant fluxes,  $\hat{T}^{f}$ is negative for 3-form fluxes, forcing the presence of orientifold planes with a negative contribution through $\hat{T}^{l}$ such that total $\hat{T}$ vanishes for a constant warping factor $A$. We shall see that for non-constant fluxes we can turn on 3-form fluxes with a non-trivial warping factor $A$  without requiring the presence of orientifold 3-planes \label{Not sure}.\\

\subsection{Flux contribution}
Let us start by studying under which conditions integration of $\hat{T}^{f}$ 
\be
\int \sqrt{g_6} e^{4A}\left(F_3^2+e^{-2\phi} H_3^2\right) d^4ydzd\bar{z},
\label{eq:Tflux}
\ee
vanishes for non-trivial 3-form fluxes. For that let us consider two simple cases:

\begin{enumerate}
\item
There is no NS-NS flux, i.e., $H_3=0$. Our strategy is to find a configuration such that the integrand in (\ref{eq:Tflux}) could be written as a holomorphic  function (plus its complex conjugate) and therefore, it would follow  $-$by the Global Residue Theorem$-$ that the integral over the punctured sphere would vanish. Notice first that
\be
F_3^2=2|\partial_z c_{ij}|^2.
\ee
By taking $\partial_zc_{ij}=\pm i \partial_z e^{-2A}$ for fixed values of $i$ and $j$ ($c_{ij}=-c_{ji}$), we see that
\begin{eqnarray}
\int \sqrt{g_6} e^{4A}F_3^2 d^6y&=&2\int \sqrt{g_6}|\partial_z A|^2 d^6y\nonumber\\
&=&{\cal V}_4 \int_{\mathbf{S}^2} \omega_1\wedge\bar{\ast}_2\omega_1,\nonumber\\
&=&\sum_p\int_{\partial{\cal U}_p}\omega_1=0,
\end{eqnarray}
with $\omega_1=g_4^{1/4}\partial_zA dz$. This happens if $d\omega_1=0$ which it is true for $\partial_zg_4$ and $\partial_zA$ being holomorphic functions on $S^2_\ast$. A trivial case involves $A=0$. Notice also the above relation among $\partial_z c_{ij}$ and $\partial_z e^{-2A}$ defines in turn a pair of holomorphic functions $U^\pm_{1, ij}$ for a fixed pair $(i,j)$ given by 
\be
U_{1,ij}^\pm=-U_{1,ji}^\pm= e^{-2A}\pm i c_{ij}.
\label{case1}
\ee
\item
There is no RR flux, i.e., $F_3=0$. By a similar procedure as before,
\be
H_3^2=2|\partial_z b_{ij}|^2,
\ee
and by taking $\partial_z b_{ij}=\pm i\partial_z e^{-2A+\phi}$ one gets
\be
\int \sqrt{g_6} e^{4A-2\phi}H_3^2 d^6y= {\cal V}_4 \sum_i\int_{\sum{\partial{{\cal U}_i}}}g_4^{1/4}\partial_z(-2A+\phi) dz,
\ee
which vanishes for $\partial_z\phi$, $\partial A$ and $\partial g_4$ being holomorphic. The trivial case happens for  $2A=\phi$. This defines a second holomorphic function
\be
U_{2,ij}=e^{-2A+\phi}\pm i b_{ij}.
\label{case2}
\ee

\item
We can now move to a more general case in which both RR and NS-NS fluxes are turned on.  The flux contribution in (\ref{eq:Tflux}) can be written as
\be
4{\cal V}_4\int e^{2D}|h(z,\bar{z})|^2 e^{2(2A+\sigma)}\left(|\partial_z c|^2+e^{-2\phi}|\partial_z b|^2\right)dzd\bar{z},
\label{Tfluxg}
\ee
where $\sqrt{g_4}=e^{2\sigma}$. By defining the same holomorphic functions $U_1$ and $U_2$ as previously, we can prove by taking
\be
\phi=2A+\sigma,
\label{case3}
\ee
that the integral (\ref{Tfluxg}) reduces to 
\be
32{\cal V}_4\int e^{2D}|h^2(z,\bar{z})|^2 |\partial_z\phi|^2 dzd\bar{z},
\ee
which vanishes for $\partial_z\phi$ being holomorphic (i.e. $\nabla^2\phi=0$), implying that the internal metric of $T^4$ must also be composed of harmonic elements.

\end{enumerate}

It is important to remark that holomorphic functions $U_{1,ij}(z)$ and $U_{2,ij}(z)$ defined by (\ref{case1}) and (\ref{case2}) determine the function $h(z,\bar{z})$ in (\ref{h}).

\subsection{Local source's contribution}
We now compute the contribution of local sources to $\hat{T}^l$ in Eq.(\ref{Tf+Tl}). Let us concentrate on $D5$-branes and the corresponding action
\be
S=\int_{{\cal W}_6}\left(-T_5 \ast_6 \mathbf{1}+\mu_5 C_6\right),
\label{eq:actionD5}
\ee
over the $D5$-brane worldvolume ${\cal W}_6$ where $T_5$ is the $D5$-brane tension. Now, since a $D5$-brane is a RR charged object in the 10-dimensional space-time, from the action
\be
S_{gauge}=\int_{X_{10}}\left(\frac{1}{2\kappa_{10}^2}F_7\wedge \ast F_7 + \mu_5 C_6\wedge PD({\cal W}_6)\right),
\ee
where $PD({\cal W}_6)$ is the Poincar\'e dual 4-form related to ${\cal W}_6$, one can compute that the $D5$-brane current $\ast J_6$ is given by
\be
\ast_{10} J_6= 2\kappa_{10}^2\mu_5 PD({\cal W}_6),
\ee
and therefore, the action (\ref{eq:actionD5}) can be written as
\be
S=\int_{X_{10}} \xi_6\wedge \ast_{10}J_6= \int \sqrt{G_{10}}\xi\cdot J,
\ee
where $\xi_6=-\frac{1}{2\kappa_{10}^2\mu_5}T_5\ast_6 \mathbf{1}$ and $\xi\cdot J= \xi_{P_1\cdots P_6}J^{P_1\cdots P_6}$. From this action, the tensor $T_{MN}^l$ is given by
\be
T_{MN}^l= \frac{1}{6!}\left(G_{MN}\xi\cdot J- 12\xi_M\cdot J_N\right),
\ee
where $\xi_M\cdot J_N=\xi_{MP_1\cdots P_5}J_N^{P_1\cdots P_5}$. Computing the internal and external parts for $D5$-branes, one gets that
\be
T_\mu^\mu=-e^{2A}T_m^m=-\frac{1}{3 \cdot 5!}\frac{e^{2A}}{\kappa_{10}^2\mu_5}\xi\cdot J,
\ee
from which $\hat{T}^l=T^\mu_\mu$. Hence the local sources contribution to the scalar curvature in (\ref{Tf+Tl}) is
\be
-\frac{1}{3\cdot 5!}\frac{1}{\kappa_{10}^2\mu_5}\int \sqrt{g_6} e^{6A} \xi\cdot J d^4 y dz d\bar z=-\frac{1}{3\cdot 5!}\int_{X_{10}}e^{6A}\xi_1\wedge\ast_{10}J_6,
\ee
Notice that D5-brane's tension is encoded in the 1-form $\xi_1$. Since $dF_3=\ast_{10} J_6$, the above integral vanishes due to Bianchi identity and by cancelation of 5-brane charges in $\mathbf{S}^2$ by the use of the GRT (see Appendix \ref{GRT}). Notice that this term also vanishes for usual compactification scenarios with constant 3-form fluxes where 5-branes are not present or wraping 2-cycles in the compact internal manifold $X_6$.

%-------------------------------------------------------------------------------------------------

\subsection{Soft terms}
Flux compactification on $T^4\times_z \mathbf{S}^2$ breaks half of the supersymmetries inherited from the ten-dimensional effective theory, while an appropriate selection of fluxes can break the rest of them  in 4-dimensions. Therefore, we expect the appearance of soft terms in the effective potential  through the interaction between moduli and four-dimensional photons. These four-dimensional photons are expected to be present as effective fields induced by the presence of D5-branes, this is, the Dirac-Born-Infeld (DBI) action of a D5-brane is effectively proportional to a DBI action of a D3-brane for a D5-brane wrapping a two-cycle in $T^4$. A detailed derivation of this result is shown in Appendix \ref{DBI5-3}. The specific form of the soft terms are deduced from the Taylor expansion of the involved fields on the effective D3-branes.\\

It is important to notice that it is expected that these D5-branes break the same supersymmetries as the 3-form RR fluxes, since they are precisely the corresponding source. This is quite the opposite with the standard flux compactification with 3-form fluxes and D3-branes, where the latter source 5-form fluxes which in principle are not related with the set of 3-form fluxes. Therefore, in that case, it is necessary to show that the bunch of D3-branes are compatible with the supersymmetry breakdown produced by the 3-form fluxes. This happens for a set of BPS D3-branes and ISD 3-form fluxes.\\

Since we only require local information around the place in which the D-branes are localized, we define the
Taylor expansion of  $X=\{\phi, A, D, \ln g_{ij}\}$ as 
\be
e^{X} = e^{X_0} +\del_z e^X z +\bar \del_z e^X \bar z+ \frac{1}{2}(\del_z \del_z e^{X} ) z^2 +  (\del_z \bar \del_{ z} e^{X}) z \bar z  + \frac{1}{2}(\bar \del_{z} \bar \del_{z} e^{X}  ) \bar z^2+ ... \, ,
\ee
where $X_0$ is the value of $X$ at a singular point on $\mathbf{S}^2$. 
Usually,  linear terms on $z$ and $\bar{z}$ must vanish in order to avoid instabilities coming from transitions among branes and fluxes \cite{Camara:2003ku}. In the usual flux compactification scenarios with constant field strengths, the constraint
\be \label{eq:lcons}
\del_z e^X z +\bar \del_z e^X \bar z = 0 \, ,
\ee
is achieved by addition of orientifold planes \cite{Camara:2003ku}. In our set up, we shall see that linear terms vanish by the appropriate selection of holomorphic functions $U$ at the singular points where the corresponding branes are localized. Notice that in the Taylor expansion
there are $z$ and $\bar z$ terms according to our anzats in which all the variables are allowed to vary
only on the compactified complex plane.\\

As shown \cite{Camara:2003ku,Camara:2004jj}  the bulk fields couple to the 4-dimensional world-volume fields of the effective $D3$-branes via DBI and CS terms.
This induces the soft term Lagrangian
\be
L_{\rm soft} = {\cal F}(z,\bar{z})\left(L^{(1)} + L^{(2)}\right) \, ,
\ee
where the function 
\be
{\cal F}(z,\bar{z})=\frac{\kappa_5}{\kappa_3} \int d^2\xi \sqrt{det ({\cal F}_{ij})} ,
\ee
as follows from Eq.(\ref{eq:DBI5-3}). The Lagrangians $L^{(1)}$ and $L^{(2)}$ are constructed from the effective DBI action of a D3-brane, and are given by
\begin{eqnarray}
L^{(1)}&=& -(m^2)^{\alpha\beta} \phi_\alpha \phi_\beta^{*} - \left( \frac{1}{3}\cA^{\alpha\beta\gamma} \phi_\alpha \phi_\beta \phi_\gamma +\frac{1}{2} \cB^{\alpha\beta}\phi_\alpha \phi_\beta -\frac{1}{2} M^\theta \lambda_\theta \lambda_\theta + {\rm h.c} \right) \, \nonumber\\
L^{(2)} &=& - \frac{1}{2}\mu^{\alpha\beta} \psi_\alpha \psi_\beta + \frac{1}{2} \cC^{\alpha\beta\gamma} \phi_\alpha \phi_\beta^* \phi_\gamma^* + M_g^{\alpha \theta} \psi_\alpha \lambda_\theta,
\end{eqnarray}
where, following notation in \cite{Camara:2003ku,Camara:2004jj}, $\phi_\alpha$ are the 3 complex scalars on the gauge 4-dimensional supersymmetric theory on the D3-branes' worldvolume while the gaugino and the 3 fermionic partners of the complex scalar fields are denoted by $\lambda$ and $\psi_\alpha$ respectively. The indices $\alpha, \beta, \gamma$, run over 1 to 3 indicating the number of fields.
The trilineal terms and gaugino mass are related with the pure $(3,0)$-form component of $G_3$ as
\bea
\cA_{\alpha\beta\gamma} &= \frac{(2 \pi)^{1/2}}{3} e^{\phi_0} \epsilon_{\alpha\beta\gamma} G_{z \omega_1 \omega_2}\, , \\
M^\theta &= \frac{e^{\phi_0/2}}{2^{3/2}} G_{z \omega_1 \omega_2} \, ,
\eea
where  $\omega_1$ and $\omega_2$ are complex coordinates on $T^4$. The $\cB$-terms depend on the quadratic expansion coefficients of the warping factor and axio-dilaton. For $C_4 = 0$ they are given by
\bea
m_{z \bar z}^2 &= 2 \partial_z\partial_{\bar{z}}A+ e^{\phi_0} \partial_z\partial_{\bar{z}}\phi \, , \nn \\
\cB_{zz} &=  \partial_z\partial_{\bar{z}}A + \frac{1}{2} e^{\phi_0} \partial_z\partial_{\bar{z}}\phi \, .
\eea
The $\mu$ and $\cC$-terms are given by
\bea
\cC^{\alpha\beta}_\gamma &= \frac{e^{\phi_0} \pi^{1/2}}{2^{1/2}} \epsilon^{\alpha\beta\delta} \left( \sigma_{\delta\gamma}-\alpha_{\bar \delta \bar \gamma}^* \right) \, , \\
\mu_{\alpha\beta} &= -\frac{e^{\phi_0/2}}{2^{3/2}}\sigma_{\alpha\beta} \, , \\
M^{\alpha \theta}_g &= \frac{e^{\phi_0/2}}{2^{5/2}}\epsilon^{\alpha\bar \beta \bar \gamma}\alpha_{\bar \beta \bar \gamma} \, 
\eea
where $\sigma_{\alpha\beta}$ and $\alpha_{\bar \alpha \bar \beta}$ following the conventions of \cite{Grana:2000jj} are defined as 
the symmetric and antisymmetric part of the $(2,1)$ and $(1,2)$ components of $G_3$,
\be \label{eq:su3dec}
G_{\alpha\beta} = \frac{1}{2}\epsilon_\beta^{\bar \gamma \bar \delta}G_{\alpha\bar \gamma \bar \delta} \,, \quad G_{\bar \alpha\bar \beta} = \frac{1}{2} \epsilon_{\bar \beta}^{\gamma\delta} G_{\bar \alpha\gamma\delta} \, .
\ee
Explicitly we have
\be
\alpha_{\alpha\beta} = \frac{1}{2}\left( G_{\alpha\beta} - G_{\beta\alpha} \right) \,, \sigma_{\alpha\beta} = \frac{1}{2} \left( G_{\alpha\beta}+G_{\beta\alpha} \right) \,,
\ee
and the same terms with bar indices, where by abuse of notation, indices $\alpha, \beta$ also refer to the complex coordinates on $T^4\times _z \mathbf{S}^2$

%------------------------------------------------------------------------------------
%-------------------------------------------------------------------------------------
\section{SUSY solutions}

As it is well known for   supersymmetric flux backgrounds, Einstein's equation is satisfied if  we demand Bianchi identity, the equation of motion for the fluxes, and supersymmetry. For a non-constant flux compactification we have found some solutions to Einstein's equations for the particular choice of an internal space given by the fibered product of $T^4$ and the punctured sphere $\mathbf{S}^2$. It is desirable to show that such solutions are compatible with those obtained by demanding Bianchi identity and supersymmetry on the flux configuration. Supersymmetry is guaranteed if the KSE are solved. A family of solutions to these equations, while fulfilling Bianchi identity, were constructed in \cite{Candelas:2014jma, Candelas:2014kma} by the use of $U$-dualities and toric geometry. In this section we shall construct two different supersymmetric solutions to the KSE compatible with the solutions to  Einstein's equations of motion found in the section 2, meaning that we shall show that it is enough to consider at most, two holomorphic functions $U$ relating warping factors with field potentials. Our construction is based in the use of $U$-dualities and we shall obtain solutions which belong to the family of solutions found and reported in \cite{Candelas:2014jma, Candelas:2014kma}. \\

Let us start with a 6-dimensional internal space given by a $T^4$ fibration over $\mathbf{S}^2$ with no fluxes. For sake of simplicity we shall take $T^4$ as the product of two identical $T^2$. Therefore, the 10-dimensional metric is given by \cite{Candelas:2014jma, Candelas:2014kma}
\be \label{eq:flat-metric}
ds^2 = \eta_{\mu \nu}dx^\mu dx^\nu+\frac{1}{v} | dy^1 + U dy^2 |^2 + \frac{1}{v} | dy^3 + U dy^4 |^2 + e^{2 D} | h(z) |^2 dz d\bar{z} \,, 
\ee
where $U$ is a holomorphic function of $\mathbb{C}$ given by
\be \label{eq:holo_1}
U (z) = v(z,\bar{z}) + {\rm i}b(z,\bar{z}) ,
\ee
playing the role of the complex structure on each $T^2$. It is well known that the above metric is solution of the 10 dimensional equations of motion in the absence of fluxes for constants $\phi$ and $D$. Thus, by T-dualizing the metric (\ref{eq:flat-metric}), part of the geometry transforms into fluxes and the six dimensional metric is modified.\\

Hence, we shall construct non-trivial solutions to the equations of motion by
performing T and S dualities. 
Then by combining these solutions we shall show that it is possible to construct new supersymmetric solutions of the 
equations of motion that are not directly related with a purely geometric background. 

%-----------------------
\subsection{Solution 1:  $H \neq 0$ and $F = 0$} \label{sec:case1}
By applying T-duality on coordinates $y^1$ and $y^3$ on  (\ref{eq:flat-metric})  we get the metric 
\be \label{eq:metric_1}
ds^2= \eta_{\mu\nu}dx^\mu dx^\nu + v\delta_{ij}dy^i dy^j+v^2|h(z)|^2 dzd\bar{z},
\ee
with the warping factors
\be
g_{ij} = v\delta_{ij} ,\; \;  A = 0 ,\; \;   e^{2 D} = v^2  \, ,
\ee
and the NS-NS potential
\be \label{eq:flux_1}
B_2= b \left( dy^1 \w dy^2 + dy^3 \w dy^4 \right), \quad \phi = D . 
\ee 
 With the purpose to solve the KSE,  as we shall see,  the imaginary
part of $U$ is related to the axion coming from the NS-NS  sector while the real part corresponds to  the saxionic partner
(thorough this work we impose $v > 0$). This follows from the fact that the gravtitino variation given in (\ref{eq:KSE-int1}) is trivially solved by $A=0$ and $F_3 = 0$. Notice that this is the trivial solution for Einstein's equation  in the case 2 shown in (\ref{case2}).\\

Expanding the covariant derivative in terms of the spinor connection (see Appendix C for notation), the internal component of  (\ref{eq:KSE-int2}) corresponding to $T^4$, reduces to
\be
\left( 2 \omega_{i \underline{i z}}\gamma^{\underline{i z}} + H_{z i {j}}\gamma^{z j} \right) \gamma^z \eta^a + \{ z \rightarrow \bar z \} = 0 \, ,
\ee
where $\{z \rightarrow \bar z \}$ stands for the same terms interchanging $z$ by $\bar z$ and underlying indices are flat indices. The above equation
is solved by (\ref{eq:gamma_eig}) which shows us that $\gamma^{z {j}}$ are nilpotent matrices for the given
metric. Similarly, for the internal component of (\ref{eq:KSE-int2})  corresponding to the compactified complex plane, we get
\bea
\left(\del_z + \frac{1}{2}\omega_{z \underline{z} \underline{\bar{z}}}\gamma^{\underline{z}\underline{\bar z}}
-\frac{(-1)^i}{4}H_{z \underline{ij}} \gamma^{\underline{ij}} \right) \eta^a  = 0 \, , 
\eea
which by direct substitution of (\ref{eq:spin_1}) and  (\ref{eq:gamma_eig})  and by choosing 
\be \label{eq:sign_eig}
\gamma^{1 2} \eta^a = \gamma^{34} \eta^a=  (-1)^{a+1} \frac{\rm i}{v} \eta^a \, ,
\ee
the gravitino equations reduce to
\be
\left( \del_z + \frac{1}{4} \frac{\del_z h}{h} \right) \eta^a = 0 \, ,
\ee
with a similar equation for $\bar z$.  Both equations 
are solved by
\be \label{eq:cond1}
\eta^a = \left(\frac{\bar{h} (\bar z)}{h (z)} \right)^{1/4} \eta_0^a \, ,
\ee
for a constant chiral spinor $\eta_0^a$. For (\ref{eq:KSE-int3}) we have then 
\be
\left( \del_z \phi - \frac{(-1)^a}{2} H_{z i j} \gamma^{i j}\right) \gamma^z \eta^a  + \{ z \rightarrow \bar{z} \} \gamma^{\bar{z}} \eta^a = 0 \, .
\ee
After substitution of (\ref{eq:metric_1}), (\ref{eq:flux_1}) and using (\ref{eq:sign_eig}) we get
\be \label{eq:sol1-dil}
\bar \del_{\bar z} \bar U \gamma^{\bar{z}} \eta^a = 0 \, , 
\ee
which vanishes for the choice
\be \label{eq:con1a}
\gamma^{\bar z} \eta^a= 0 \, .
\ee
This condition implies that half of the components of the chiral fermion are annihilated, 
and hence the solution preserves half of the original supersymmetries, this is ${\cal N}=4$ in four dimensions. Notice that by (\ref{eq:con1a})  the two internal spinors
are orthogonal. At this point, we can notice that the complex conjugate brings an additional solution by choosing
\be
\gamma^{z}\eta^a = 0 ,\quad \gamma^{12} \eta^a = \gamma^{3 4} \eta^a =  (-1)^{a} \frac{\rm i}{v} \eta^a \, ,
\ee
and
\be \label{eq:cond1}
\eta^a = \left(\frac{h (z)}{\bar{h} (\bar z)} \right)^{1/4} \eta_0^a \, .
\ee
This configuration corresponds to  a SU(2) structure solution as the one shown in \cite{Candelas:2014jma,Candelas:2014kma} with
vanishing  $C_0$ and  $C_2$. The holomorphic three-form $\Omega$ and the K\"ahler form $J$ are given by
\bea
\Omega =  h(z) v^2 dz \w (dy^2 + {\rm i} dy^1) \w (dy^4 + {\rm i} dy^3) \, , \nn \\
J = v \left( dy^1 \w dy^2 + dy^3 \w dy^4 \right) +\frac{\rm i}{2} |h(z)|^2 v^2 dz \w d\bar z \, .
\eea
Since $dJ\neq 0$ and $d\Omega\neq 0$ the internal six-dimensional space is a complex manifold with vanishing torsion classes\footnote{The globally defined $J$ and $\Omega$ forms are decomposed in terms of the invariant
$j$ and $\Omega_2$ forms according with its SU(2) structure. This choice determines uniquely
the complex structure for a given metric \cite{Hitchin:2000sk,Larfors:2010wb}.} $\cW_1 = \cW_2 = \cW_4 = 0$, corresponding to a non-K\"ahler warped complex manifold.\\

This
choice of $\Omega$ and $J$ automatically defines the complex structure with holomorphic coordinates,
$dw_1 = dy^2 + {\rm i} dy^1$ and $dw_2= dy^4 +{\rm i} dy^3$. Notice that the complex structure for
this case is a positive imaginary number. Notice also that the superpotential
\be
W=\int G_3\wedge \Omega=-i \int e^{-\phi}H_3\wedge \Omega,
\ee
as well as the K\"ahler derivatives with respect to the dilaton and complex structure moduli, vanish at the minimum implying that we actually have constructed a ${\cal N}=4$ SUSY Minkowski vacuum.

\subsubsection{Soft terms for $H \neq 0$ and $F = 0$}
Given that half of the supersymmetry is broken in the above setup, it is possible to compute the corresponding induced soft terms on the effective 4-dimensional theory derived from $D5$-branes. The vanishing of the linear term (\ref{eq:lcons})  puts a restriction on the
holomorphic function $U(z)$. Since the holomorphic function is only allowed to vary in the base
it is convenient to split the complex coordinates $z$ in real and imaginary components as,
\be
dz = dy^5 + {\rm i}dy^6 \, ,
\ee
for which  (\ref{eq:lcons}) reads\footnote{We employ
the definition of holomorphic derivative compatible with its complex structure, 
namely, $\del_z = \del_{y^5} + \tau \del_{y^6}$, with $\tau$ the complex structure in $\mathbf{S}^2.$}
\be
\del_{5} v = 0 \, ,
\ee
implying that the saxionic components are only allowed to vary along $y^6$. 
According to (\ref{eq:su3dec}) we find that
\be
\sigma_{ab}  = 0 \, \quad \alpha_{\omega_1 \omega_2} = \alpha_{\bar \omega_1 \bar \omega_2}^* = -{\rm i}\frac{\bar \del_{\bar z} \bar U}{2 v} \, ,
\ee
where the mixing of $\alpha_{\omega_1 \omega_2}$ and $\alpha_{\bar \omega_1 \bar \omega_2}$ implies that
the flux configuration is composed of ISD and IASD fluxes in $T^4\times _z\mathbf{S}^2$. The vanishing soft terms are then given by
\be
\cA_{z \omega_1 \omega_2} = M^\theta = \mu_{\alpha\beta} = 0 \, ,
\ee
and
\bea
\cB_{zz} &= \frac{(\del_z U)^2 - v \del_z^2 U}{2 v^2}  \, , \nn m^2_{z \bar z} = \frac{1}{2 v^2} |\del_z U|^2  \, , \\
\cC^{z \omega_1}_{\omega_1} &= \cC^{z \omega_2}_{\omega_2}= -{\rm i}\frac{(\pi)^{1/2}}{2^{3/2}}\bar \del_{\bar z} \bar U \, , M^{a\theta}_g = -{\rm i}\frac{1}{2^{5/2}v^{1/2} } \del_z U \, .
\eea
It is convenient to express all soft terms as functions of the internal space volume. 
In particular, since the function $h(z)$ is absent in all the above expressions, the soft terms only depend
on the volume of the fiber in the Einstein frame defined as\\
\be
\hat\cV_4 = \int_{T^4} dy^4 \sqrt{g_4^E} \,.
\ee 
The  soft terms are given by
\be
\cB_{zz} = -\frac{1}{2}\frac{\del^2_z U}{\hat\cV_4}+\frac{1}{2}\frac{(\del_z U)^2}{\hat\cV_4^2} \,, m^2_{z\bar z}=\frac{1}{2}\frac{|\del_z U|^2}{\hat\cV_4^2} \,,
\ee
and 
\be
M^{a\theta}_g = -\frac{{\rm i}}{2^{5/2}} \frac{\del_z U}{\hat\cV^{1/2}_4} \,.
\ee
Thus at leading order the soft masses depends inversely  on the volume of the compact space, which
is compatible with soft masses arising in the large volume scenario \cite{Blumenhagen:2009gk}.\\

Since both,  $(3,0)$ and $(0,3)$ components of $G_3$ are zero, this case corresponds to a non-scale supersymmetric
vacuum with zero cosmological constant\footnote{From a phenomenological point of view, the absence of $\mu$-terms may
potentially lead to problems in the CMSSM by generating the right electroweak scale ($\mu$-problem). However,
since only two coordinates are allowed to vary, these results are compatible with universal masses for the
squarks and sleptons with vanishing gaugino mass.}.\\

%-----------------------
\subsection{Solution 2: $H = 0$ and $F \neq 0$} \label{sec:case2}

Consider now the same metric (\ref{eq:metric_1}) with a NS-NS potential $C_2$ given by
\be
C_2=c(z,\bar{z})(dy^2\wedge dy^3+dy^1\wedge dy^4).
\label{eq:flux_2}
\ee
The S-dual configuration is then given by the metric (\ref{eq:manzats}) with the parameters 
\be \label{eq:metric_2}
g_{ij} = {\rm diag } \left( 1 , 1, 1 , 1 \right) ,\; \;  e^{2A} = \frac{1}{v}  ,\; \;   e^{2 D} = v \, ,
\ee
with a RR field potential $C_2$ and $\phi = 2 A$. This flux and metric configuration also leads to ${\cal N}=4$ in four dimensions. However,  the relative orientation of the internal spinors changes,  implying 
different solutions for the Killing equations. It is important to notice that this solution, obtained by  S-duality, corresponds to the trivial solution $A=0$ corresponding to the holomorphic function $U_1$ shown in (\ref{case1}) which fulfils  Einstein's equations. It seems that S-duality relates different solutions of Einstein's equations. As in the previous solutions, we also require a single holomorphic function $U=e^{-2A}+\rm{i}c$.\\

A direct substitution of (\ref{eq:metric_2}) and (\ref{eq:flux_2}) on the first Killing spinor equation  (\ref{eq:KSE-int1}) leads to
\be \label{eq:ccon2}
\del U \gamma^z \left( \eta^1_+ + \eta^2_+ \right) + \bar \del_{\bar z} \bar U \gamma^{\bar z}\left( \eta^1_+ - \eta^2_+ \right)  = 0 \, ,
\ee
which is solved by taking
\be \label{eq:con2}
\eta^1_+ = -\eta^2_+ ,\quad \gamma^{\bar z} \eta^1_+  = 0 \, ,
\ee
or
\be \label{eq:cond2a}
\eta^1_+ = \eta^2_+ ,\quad \gamma^{ z} \eta^1_+  = 0 \, ,
\ee
for which we have used (\ref{eq:gamma_eig}) and 
\be \label{eq:sign_eig2}
\gamma^{1 4} \eta^a = \gamma^{2 3} \eta^a = (-1)^a {\rm i} \eta^a \, .
\ee
The second  set of Killing spinor equations (\ref{eq:KSE-int2}) decomposes on two equations corresponding to $T^4$ and $\mathbf{S}^2$. For the toroidal contribution,  the
gravitino variation reduces to
\be
2\omega_{i \underline{i z}} \gamma^{i \underline{i z}} + \left(
3 F_{i j z} \gamma^{j z}  - F_{j k z} \delta^{j}_{i} \gamma^{k z} \right) \eta^a + \{ z \rightarrow \bar{z} \} = 0 \, ,
\ee
which vanishes by relabelling the dummy indices and by the nilpotency of the gamma matrices (\ref{eq:gamma_eig}). For the complex plane contribution, all contractions of the form $\gamma^{i j \bar z}\gamma_z $ vanish and thus
the gravitino variation reduces to
\be
\left( \del_z + \frac{1}{2}\omega_{z \underline{z}\underline{\bar z}}\gamma^{\underline{z\bar z}} \right) \eta^a +\frac{(-1)^i}{8}e^{\phi} \left( F_{2 3 z}\gamma^{2 3}+F_{1 4 z}\gamma^{1 4} \right) \tilde \eta^a= 0.
\ee
By substitution of (\ref{eq:spin_1}) and (\ref{eq:con2}) in  (\ref{eq:sign_eig2}) we get
\be \label{eq:KSE-z2}
\left( \del_z + \frac{1}{4} \frac{\del_z h}{h} + \frac{1}{4} \frac{\del_z U}{v} \right) \eta^a = 0 \, ,
\ee
with a similar equation for the complex conjugate coordinate $\bar{z}$. The set of coupled differential
equations are solved by
\be \label{eq:KSE-sol2b}
\eta^a = \left( \frac{\bar h(\bar z) }{h (z) v}\right)^{1/4} \eta^a_0 \, .
\ee
From equation (\ref{eq:KSE-int3}) we realize that for $\phi = 2 A$ and $H = 0$, the dilatino variations
and (\ref{eq:KSE-int1}) have the same form and are both satisfied fulfilled by (\ref{eq:con2}). We notice that this configuration is a
special case of a SU(2) structure solution presented in \cite{Candelas:2014jma} with
\bea
\Omega =  h(z) v^{1/2} dz \w (dy^4 + {\rm i} dy^1) \w (dy^3 + {\rm i} dy^2) \, , \nn \\
J =  \left( dy^2 \w dy^3 + dy^1 \w dy^4 \right) +\frac{\rm i}{2} |h(z)|^2 v dz \w d\bar z \, ,
\eea
and vanishing torsion clases $\cW_1 = \cW_2=\cW_3=\cW_4=0$  corresponding to a K\"ahler manifold with
holomorphic coordinates $dw_{1}= dy^4 + {\rm i} dy^1$ and $dw_{2}=dy^3 + {\rm i} dy^2$ with a positive and imaginary complex structure. As in the previous case, the superpotential and all K\"ahler derivatives vanish at the minimum showing that we have constructed a ${\cal N}=4$ SUSY Minkowski vacuum. This is confirmed by studying the generated soft terms.

\subsubsection{Soft terms for $H = 0$ and $F \neq 0$}
By direct substitution of (\ref{eq:metric_1}) and (\ref{eq:flux_1}) into (\ref{eq:lcons}) we realize
that the linear term corresponding to the warping factor $v$  and  the dilaton vanish for $\del_{5} v = 0$.
The associated soft terms $\sigma_{\alpha\beta}$ and $\alpha_{\alpha\beta}$  for the ISD and AISD components of $G_3$ are then given by
\be
\sigma_{\alpha\beta} = 0\,, \quad \alpha_{w_1 w_2} = \alpha_{w_1 w_2}^* = \frac{\del_z U}{2} \, .
\ee
The vanishing soft terms are 
\be
\cA_{z w_1 w_2}= M^\theta = \mu_{\alpha\beta} = 0 \, ,
\ee
while the scalar masses and the $\cB$ term are given by
\be
m^2_{z\bar z} = \frac{1}{2 v^3}|\del_z U|^2 \,, \quad \cB_{zz} = \frac{(\del_z U)^2 +v(v - 1)\del_z^2 U}{2 v^3} \, .
\ee
Similarly, the $\cC$ and $M^{aI}_g$ terms are
\be
\cC^{z w_1}_{w_1} = \cC^{z w_2}_{w_2}= \frac{(\pi)^{1/2}}{2^{3/2}}\bar \del_{\bar z} \bar U \, , M^{a\theta}_g =\frac{1}{2^{5/2}v^{1/2} } \del_z U \, .
\ee
Notice that as in the previous case, the trilineal terms vanish due to the absence of $(3,0)$-forms.
Also this pattern of soft terms is compatible with a non-scale supersymmetric
vacuum with vanishing cosmological constant.\\

On the other hand, in the Einstein frame,  the non-vanishing soft terms  suppressed by the internal volume are
\be
\cB_{zz} = \frac{1}{2}\frac{\del_z^2 U}{\hat\cV_4}-\frac{1}{2}\frac{\del_z^2 U}{\hat\cV_4^2}+\frac{1}{2}\frac{(\del_z U)^2}{\hat\cV_4^3}\,, m^2_{z\bar z} = \frac{1}{2}\frac{|\del_z U|^2}{\hat\cV^3_4} \,,
\ee
and
\be
M^{a\theta}_g = \frac{1}{2^{5/2}}\frac{\del_z U}{\hat\cV^{1/2}_4} \, .
\ee

%-----------------------
\subsection{SUSY solution with $H \neq 0$ and $F \neq 0$}\label{sec:case3}
In this section we propose a non-trivial supersymmetric flux configuration by   combining  the solutions showed in section \ref{sec:case1} and \ref{sec:case2}. Specifically we propose an internal metric related to the presence of RR and NS-NS potentials supported on 3-cycles such that $F_3\wedge H_3=0$ with a metric given by (\ref{eq:manzats}) with conformal factors
\be \label{eq:metric_3} 
g_{ij} = {\rm diag} \left( v_2 , v_2, v_2, v_2 \right) , \; e^{2 A} = \frac{1}{v_1} ,\; e^{2D} = v_1 v_2^2 \, , 
\ee
and potential fluxes
\be \label{eq:flux_3}
C_2 = c \left( dy^2 \w dy^3 + dy^1 \w dy^4 \right) ,\quad B_2 = b \left( dy^1 \w dy^2 + dy^3 \w dy^4 \right) ,\quad e^{\phi} =  \frac{v_2}{v_1} \, .
\ee
Notice that these potentials are both self-dual (SD) in $T^4$ and correspond to a solution of Einstein equations given by the family (\ref{case3}). 
For this case we require two holomorphic functions $U_1$ and $U_2$ given by the trivial solutions constructed above.
Observe  that Eq. (\ref{eq:KSE-int1}) does not depend on the NS fluxes, and in consequence can be similarly reduced  as (\ref{eq:ccon2}). 
Thus the choice
\be \label{eq:con3}
\eta^1_+ = -\eta^2_+ ,\quad \gamma^{\bar z} \eta^1 = 0 \, ,
\ee
or
\be \label{eq:con3a}
\eta^1_+ = \eta^2_+ ,\quad \gamma^{z} \eta^1 = 0 \, ,
\ee
solves the first Killing equations  (\ref{eq:KSE-int1}). Hereafter we shall use condition (\ref{eq:con3}) to solve the rest of the Killing spinor equations but a similar result is obtained by considering the second condition (\ref{eq:con3a}). These solutions establishes a ${\cal N}=2$ SUSY four-dimensional theory.\\

Hence, the toroidal internal component of (\ref{eq:KSE-int2})  reduces to
\be \label{eq:nil3}
2\left( \omega_{i \underline{j z}} \gamma^{\underline{j z}} +  H_{i j z} \gamma^{j z} \right) \eta^a + e^{\phi} \left(
F_{i j z} \gamma^{j z}  - F_{j k z} \delta^{j}_{i} \gamma^{k z} \right) \tilde \eta^a+ \{ z \rightarrow \bar z \} = 0 \, ,
\ee
which is satisfied by the nilpotency of $\gamma^{j z}$ (see \ref{eq:gamma_eig}). For $m = \{ z , \bar z \}$, the relevant eigenvalues for the
gamma matrices are
\be \label{eq:eval3}
\gamma^{k l} \eta^a = (-1)^{i} \frac{{\rm i}}{v_2} \eta^a \, ,
\ee
for any $k \neq j$. Using these eigenvalues and (\ref{eq:con3}), the gravitino variation reduces to
\be \label{eq:KSE-z3}
\left( \del_z + \frac{1}{4} \frac{\del_z h}{h} + \frac{1}{4} \frac{\del_z U_1}{v_1} \right) \eta^a = 0 \, .
\ee
For arbitrary $v_1$,$v_2$ we have
\be \label{eq:sol3}
\eta^a = \left( \frac{\bar h(\bar z) }{h (z)v_1}\right)^{1/4} \eta^a_0 \, ,
\ee
which matches with (\ref{eq:KSE-sol2b}). Since this background satisfies both KSE and Bianchi identities
is in consequence a solution of the equations of motion. For the second solution in (\ref{eq:con3a}),  the corresponding solution to  gravitino equation is given by the complex conjugate of (\ref{eq:sol3}). In this case
the holomorphic 3-form and K\"ahler form are given by
\bea
\Omega =  h(z) v_1^{1/2} v_2^2 dz \w (dy^2 - {\rm i} dy^1) \w (dy^4 + {\rm i} dy^3) \, , \nn \\
J =  v_2 \left( dy^1 \w dy^2 + dy^3 \w dy^4 \right) +\frac{\rm i}{2} |h(z)|^2 v_1 v_2^2 dz \w d\bar z \, .
\eea
This corresponds to a manifold with vanishing torsion clases $\cW_1 = \cW_2 = \cW_4 = 0$ and holomorphic coordinates $dw_1 = dy^2 - {\rm i}dy^1$ and
$dw_2 = dy^4 + {\rm i}dy^3$. The complex structure is again purely imaginary. Since $H_3\wedge F_3=0$ we also have a  vanishing superpotential ${\cal W}$ leading to a Minkowski ${\cal N}=2$ SUSY vacuum.\\

The existence of these solutions can be anticipated by 
comparison of (\ref{eq:cond1}) and (\ref{eq:con2}). On one hand, these conditions are compatible in the
sense that both solutions are annihilated for the same $\gamma$ matrix. On the other hand the contributions of the NS and RR
sectors add linearly. However, by changing the relative sign of the two cycle
supported by either $C_2$ or $B_2$ these conditions change, and  all  internal components of the
internal spinors are annihilated breaking SUSY. The 4-cycle in $T^4$ the RR and NS-NS fluxes are supported on, represent a linear combination compatible with SUSY preservation\footnote{In the following
we always refer to ASD and SD with respect to $T^4$. IASD or ISD are referred  to $T^4 \times _z \mathbf{S}_\ast^2$.}. Thus by keeping either ASD or SD cycles
for both sectors the solution of the equations of motion is still supersymmetric. As we shall see in the next section, an independent linear combination of these cycles will leads us to a non SUSY configuration. \\

\subsubsection{Soft terms for $H \neq 0$ and $F \neq 0$}

When both fluxes are turned on, the linear terms in the expansion (\ref{eq:lcons}) vanishes by
imposing
\be
\del_{5} v_1 = \del_{5} v_2  = 0 \,.
\ee
Thus the real part of the two holomorphic functions are restricted to vary on one direction
of the base. For this case, we realize that the symmetric and antisymmetric parts of (\ref{eq:su3dec}) are now,
\be
\sigma_{\bar w_1 \bar w_1} = -\sigma_{\bar w_2 \bar w_2} = \sigma_{w_1 w_1}^*  = -\sigma_{w_2 w_2}^* = \frac{\del_z U_1}{2} \, , \\
\ee
\be
\sigma_{\bar w_1 \bar w_2} = -\frac{{\rm i}}{2}\frac{v_1}{v_2} \del_z U_2 \, , \quad \sigma_{w_1 w_2} = -\frac{{\rm i}}{2}\frac{v_1}{v_2} \bar \del_{\bar z} U_2 \, ,
\ee
thus, the trilineal term is now absent and the gaugino mass is zero, this is
\be
\cA_{z w_1 w_2} = M^a = M^{a\theta}_g = 0 \, .
\ee
For the scalar masses $m^{2}_{ij}$ and $\cB$-term, we have 
\bea \label{eq:mBsoft}
m^2_{z\bar z} &= \frac{1}{2 v_1^3 v^2_2}\left( { v_2^2 |\del_z U_1 |^2 + v_1^3 |\del_z U_2 |^2 - v_1^2 v_2 \del_z \bar \del_{\bar z}{\rm Re}(\bar U_1  U_2) }\right) \, , \\
\cB_{z z} &= \frac{1}{2 v_1^3 v_2^2} \left( v_1^3 (\del_z U_2)^2 +v_2^2 \left[ (\del_z U_1)^2+(v_1^2-v_1)\del_z^2 U_1 \right] - v_1^2 v_2 \left[ \del_z U_1 \del_{z} U_2 +v_1 \del_z^2 U_2\right]\right) \, ,
\eea
and the $\cC$ and $\mu$ terms are
\be
\cC^{z w_1}_{w_1} = -\cC^{z w_2}_{w_2}= -{\rm i}\frac{\pi^{1/2}}{2^{3/2}}\bar \del_{\bar z} \bar U_2 \; , \cC^{z w_1}_{w_2} = \cC^{z w_2}_{w_1} = \frac{\pi^{1/2}}{2^{3/2}} \frac{v_2}{v_1}\bar \del_{\bar z} \bar U_1 \, ,
\ee
\be
\mu_{w_1 w_1} = -\mu_{w_2 w_2} = -\frac{1}{2^{5/2}}\frac{v_2^{1/2}}{v_1^{1/2}}\bar \del_{\bar z}\bar U_1 \,, \mu_{w_1 w_2} = \mu_{w_2 w_1} = \frac{{\rm i}}{2^{5/2}}\frac{v_1^{1/2}}{v_2^{1/2}}\bar \del_{\bar z}\bar U_2 \,, 
\ee
which is compatible with a non-scale SUSY minima. Notice that the soft terms of the previous cases are recovered by set either $U_1$ or $U_2$ to a constant. 
In the Einstein frame the the $m^2$ and $\cB$ terms are
\bea \label{eq:sfermions}
m_{z \bar z}^2 &= -\frac{\del_z \bar \del_{\bar z} {\rm Re}(\bar U_1 U_2)}{\hat\cV_4}+\frac{v_1^2 |\del_z U_2|^2}{2 \hat\cV_4^2}+\frac{v_2^3 |\del_z U_1|^2}{2 \hat\cV_4^3} \,, \nn \\
\cB_{zz} &= -\frac{\del_z U_1 \del_{z} U_2 +v_1 \del_z^2 U_2-v_2 \del_z^2 U_1}{2 \hat\cV_4}-\frac{v_2^2 \del_z^2 U_1}{2 \hat\cV_4^2}+\frac{v_1^2 (\del_z U_2)^2}{2 \hat\cV_4^2}+\frac{v_2^3 (\del_z U_1)^2}{2 \hat\cV^3_4} \,,
\eea
and the $\cC$ and $\mu$ terms
\bea \label{eq:soft1}
C_{w_1}^{z w_1} = -{\rm i}\frac{\pi^{1/2}}{2^{3/2}}\bar \del_{\bar z} \bar U_2 \,, \cC_{w_2}^{z w_1} = \frac{\pi^{1/2}}{2^{3/2}} \frac{v_2^2}{\hat\cV_4}\bar \del_{\bar z} \bar U_1 \,, \nn \\
\mu_{w_1 w_1} = -\frac{1}{2^{5/2}}\frac{v_2 \bar \del_{\bar z}\bar U_1}{\hat\cV_4^{1/2}} \,, \mu_{w_1 w_2} = \frac{{\rm i}}{2^{5/2}}\frac{v_1 \bar \del_{\bar z}\bar U_2}{\hat\cV_4^{1/2}} \,. 
\eea
From (\ref{eq:sfermions}) we realize that at leading order on $\hat\cV^4$ the squarks and sleptons becomes tachyonic with positive contributions at subleading
order. In the other hand the absence of trilinear terms as well of gaugino mass, is related to the absence of $(3,0)$ and $(0,3)$-forms. This 
result is compatible with a non-scale SUSY vacuum (vanishing vev's of the auxiliary fields for the axio-dilaton and K\"ahler moduli).

%-----------------------------------------------------------------------------------------NON-SUSY--------------------------------------------------------------------------------------------------------
%-----------------------
\section{Non-SUSY case} \label{sec:case4}
By properly choosing the orientation of the two cycles the field potentials $C_2$ or $B_2$ are supported on, we argue that SUSY shall be
broken while still a solution of the equations of motion. Let us consider the case of the metric (\ref{eq:manzats}) with conformal factors
\be \label{eq:metric_3}
g_{ij} = v_2\delta_{ij} , \; e^{2A} = \frac{1}{v_1} ,\; e^{2D} = v_1 v_2^2\, , 
\ee
and potential fluxes
\be \label{eq:flux_3}
C_2 = c \left( dy^2 \w dy^3 - dy^1 \w dy^4 \right) ,\quad B_2 = b \left( dy^1 \w dy^2 + dy^3 \w dy^4 \right) ,\quad e^\phi =  \frac{v_2}{v_1} \, .
\ee
Notice that the RR potential is ASD contrary to the corresponding potential in the previous SUSY solution.
This condition together with
(\ref{eq:con3}) implies that
\be \label{eq:prim}
\slashed{F} \eta^a = 0 \, ,
\ee
thus by (\ref{eq:KSE-int1}) we have that
\be
\del_z U_1\gamma^z \eta^a + \bar \del_{\bar z}U_1\gamma^{\bar z} \tilde \eta^a = 0 \, ,
\ee
which is satisfied by
\be
\gamma^z \eta^a_+ = \gamma^{\bar z} \eta^a_+ = 0 \, ,
\ee
implying that all internal spinors vanish and  SUSY is broken. Some
comments are in order: 
Eq. (\ref{eq:prim}) implies that $F_3$ is a primitive 3-form \footnote{The primitive condition
implies that $g^{n \bar p}F_{m n \bar p} = 0$ and similarly for $H_3$.}  which forces $H_3$ to be a  primitive as well by solving the KSE. This statement is translated to the condition
that the KSE shall be satisfied if both  NS and RR fluxes are either SD or ASD. \\

Now let us prove that the above solution is indeed a solution of the equations of motion. In absence of $F_1$
and $F_5$ fluxes, the respective equations of motion are trivially solved . Einstein equation
(in the Einstein frame) reads
\be
G_{MN}^{e} = T^{\phi}_{MN} + T^{{\rm f}}_{MN} \, ,
\ee
where 
\bea
G_{MN}^{e} &= R_{MN} - \frac{1}{2}g_{MN}R \, , \nn \\
T_{MN}^{\phi} &= \frac{1}{2} \left( \del_M \phi \del_N \phi   - \frac{1}{2}g_{MN}(\del \phi)^2  \right)  \, , \nn 
\eea
and the flux contribution to the energy stress tensor is given by
\be
T_{MN}^{\rm f} = \frac{e^{\phi}}{4}\left( F_{M IJ}F_N^{IJ} - \frac{1}{6}g_{MN}F^2  \right) + \frac{e^{-\phi}}{4}\left( H_{M IJ}H_N^{IJ} - \frac{1}{6}g_{MN}H^2  \right) \, .\nn
\ee
The four-dimensional component of the Einstein tensor reads
\be \label{eq:einten}
G_{\mu \nu}^{e} = \frac{1}{2^3 |h(z)|^2(v_1 v_2)^4} \left(  | v_2 \del_z U_1 + v_1 \del_z U_2|^2 + 2 v_2^2 |\del U_1|^2 +2 v_1^2 |\del U_2|^2 \right)\eta_{\mu \nu} \, ,
\ee
while the dilaton and flux contributions to the four-dimensional energy-stress  tensor are
\begin{eqnarray}
T_{\mu \nu}^{{\rm \phi}} &=& \frac{|v_1 \del_z U_2 + v_2 \del_z U_1 |^2}{2^3 |h(z)|^2(v_1^4 v_2^4)}\eta_{\mu \nu} \, ,\nonumber\\
T_{\mu \nu}^{{\rm f}} &=& \frac{v_2^2|\del U_1 |^2 + v_1^2 |\del U_2|^2}{2^2 |h(z)|^2(v_1^4 v_2^4)}\eta_{\mu \nu} \,,
\end{eqnarray}
showing that our configuration is an exact solution to four-dimensional Einstein equations. For the
internal components we see that the only non-zero components are
\be
G_{zz}^{e} = (G_{\bar z \bar z}^{e})^* = -\frac{\left( v_2 \del_z U_1 + v_1 \del_z U_2 \right)^2}{2^{3}v_1^2 v_2^2} \, ,
\ee
which cancels out with the dilaton and flux contribution to the internal energy-stress tensor. The dilaton equation of motion is
\be 
\frac{1}{\sqrt{g}}\del_M \left( \sqrt{g} g^{MN} \del_N \phi\right) = \frac{1}{12} \left( e^{\phi} F^2  - e^{-\phi} H^2 \right) \, ,
\ee
notice that the flux contribution only involves quadratic terms. Because of this, the change from SD to ASD
in the cycles supporting the RR and NS fluxes are still a solution of the equations of motion but not
of the KSE. A straightforward calculation shows us that the flux contribution is
\be
\frac{1}{12}\left( e^{\phi} F^2 - e^{-\phi} H^2 \right) = \frac{v_2^2 \del_z \bar \del_{\bar z} |U_1|^2 +v_1^2 \del_z \bar \del_{\bar z} |U_2|^2}{2 |h(z)|^2 v_1^{7/2}v_2^{7/2}} \, ,
\ee
which exactly cancels out with the dilaton contribution.

%-----------------------

\subsection{Soft terms for the non-SUSY case}
For the case of section \ref{sec:case4}, we have that $\alpha = 0$  and $\sigma$ has non-zero components. This
can be seen as the $G_3$ decomposes as a sum of the terms coming from the NS and RR sector given
in sections \ref{sec:case1} and \ref{sec:case2} respectively. Explicitly,
\be
\sigma_{z z} = -\sigma_{\bar z \bar z}^* =  \frac{\del U_1}{2} \,, \sigma_{w_1 w_2} =-{\rm i}\frac{v_1}{2 v_2}\bar \del_{\bar z}\bar U_2 \, ,
\sigma_{\bar w_1 \bar w_2} =-{\rm i}\frac{v_1}{2 v_2}\bar \del_{z}\bar U_2 \, ,
\ee
For this case there the non-vanishing soft terms are,
\be \label{eq:Msoft}
M^{aI}_g = 0\, .
\ee
while the trilineal term and gaugino mass are given by
\be
\cA_{z w_1 w_2} = -\frac{\pi^{1/2}}{2^{1/2}\cdot 3}\frac{v_2}{v_1}\del_z U_1 \,, \quad M^I = -\frac{1}{2^{5/2}}\frac{v_2^{1/2}}{v_1^{1/2}}\del_z U_1 \, .
\ee
The scalar masses and $\cB$ terms are given by (\ref{eq:mBsoft}). The $\mu$ and $\cC$ terms are
\bea
\cC^{z w_1}_{w_1} &= -\cC^{z w_2}_{w_2} = -{\rm i}\frac{\pi^{1/2}}{2^{3/2}}\bar \del_{\bar z}\bar U_2 \,, \cC^{w_1 w_2}_z = -\frac{\pi^{1/2}}{2^{3/2}} \frac{v_2}{v_1}\del_z U_1 \, , \\
\mu_{z z} &= -\frac{1}{2^{5/2}} \frac{v_2^{1/2}}{v_1^{1/2}}\del_z U_1 \, ,
\mu_{w_1 w_2} = {\rm i}\frac{1}{2^{5/2}}\frac{v_1^{1/2}}{v_2^{1/2}}\bar \del_{\bar z} \bar U_2 \,,
\eea
For this case, we have non-zero $(3,0)$-form, which leads us to SUSY breaking with non-zero
auxiliary field for the axio-dilaton and K\"ahler moduli. The trilinear term and gaugino mass in
the Einstein frame are rewritten as
\be
\cA_{z w_1 w_2} = -\frac{\pi^{1/2}}{2^{1/2}\cdot 3}\frac{v_2^2}{\hat\cV_4}\del_z U_1 \,, \quad M^\theta = -\frac{1}{2^{5/2}}\frac{v_2}{\hat\cV_4^{1/2}}\del_z U_1 \, ,
\ee
The squarks/sleptons massses, $\mu$ and $\cB$ terms have the same dependence on the internal volume $\hat\cV$ as in (\ref{eq:sfermions}) and (\ref{eq:soft1}).

%------------------------------

\subsection{Stability analysis}
In this part we want to study stability of the metric (\ref{eq:metric_3}). For that we shall consider a simple but illustrative case in which we are going to take into account only spherical symmetric perturbations of the warping factors on the four-dimensional space-time. This is
\bea
e^{2A} &\rightarrow e^{2A} + \epsilon \delta A(t,r) \,,\\ \nn
e^{2D} &\rightarrow e^{2D} + \epsilon \delta D(t,r) \,,\\ \nn
g_{ii} &\rightarrow g_{ii} + \epsilon \delta g(t,r) \,, \\ \nn
\phi &\rightarrow \phi + \epsilon \delta \phi(t,r)
\eea
where $r$ is a radial coordinate in the four-dimensional space and $\epsilon$ is the constant perturbative parameter. In the following we shall omit the radial and time dependence of the variations. The linearized four-dimensional Einstein equations are solved by the relations
\bea
 \label{eq:var1}
 e^{-2D}\nabla^2 \text{$\delta $D}-e^{-2 A}\nabla^2 \text{$\delta $A} +f_1 \text{$\delta $A}=0 \,, \nn\\
 \frac{1}{g_{ii}}\nabla^2 \text{$\delta $g}-e^{-2 A}\nabla^2 \text{$\delta $A}+f_2 \text{$\delta $A}=0 \,,\nn \\
 \nabla^2 \delta \phi +2 e^{-2A}\nabla^2 \text{$\delta $A}+f_3 \text{$\delta $A}=0 \,,
\eea
together with,
\bea \label{eq:var2}
\partial_{t,r} \left( e^{-2D}\text{$\delta $D}+e^{-2A} \text{$\delta $A}+\frac{2}{g_{ii}} \text{$\delta $g}+2 \delta \phi \right)=0 \,, \nn \\
 \partial_{t}\left(3 e^{-2D}\text{$\delta $D}+\frac{1}{g_{ii}}\text{$\delta $g}\right)=0 \,, \nn \\
 \partial_{t}\left( 3 e^{-2A} \text{$\delta $A}+2 \delta \phi+\frac{1}{g_{ii}}\text{$\delta $g} \right)=0 \,,
\eea
and the relations among the variations
\be \label{eq:var3}
 f_4 e^{-2A} \text{$\delta $A}=\frac{e^{-2D}}{16}\text{$\delta $D} \,, \quad f_5 e^{-2A} \text{$\delta $A}=\frac{1}{4 g_{ii}}\text{$\delta $g} \,,
\ee
where $f_\ast \left(z , \bar{z} \right)$, with $\ast=\{1,2,3,4\}$ are functions  on $z$ and $\bar{z}$ \footnote{The explicit form of $f_\ast$, although relevant for the following discussion, is omitted since it is a quite intricate function of $z$ and $\bar{z}$.}. Combining (\ref{eq:var1}) and (\ref{eq:var3}), we obtain  
\be
\nabla^2 \delta X + \cF_\ast \delta X = 0  \,,
\ee
where $\delta X = \{ A, D, g \}$ and $\cF_\ast \equiv \cF_\ast (z, \bar z)$. For $\delta \phi$, we notice that,
\bea
3 e^{-2 A} \delta A + 2 \delta \phi + \frac{1}{g_{ii}} \delta g = \cM(r,z,\bar z) \,, \nn \\
3 e^{-2 D} \delta D + \frac{1}{g_{ii}} \delta g = \cN (r,z, \bar z) \,, 
\eea
which together with (\ref{eq:var1}), gives
\be
\bigg( 1 + \frac{2}{3}\frac{1}{16 f_4 -1}\bigg)\nabla^2 \delta \phi +	\frac{2}{3}\frac{f_3 e^{2A}}{16 f_4-1}\delta \phi  +
2 \left( \frac{\nabla^2 \cM - \nabla^2 \cN}{3(1-16 f_4)}\right) - \frac{\cM - \cN}{3(16 f_4-1)} f_3 e^{2A} = 0 \,.
\ee
For $\cM = \cN$ the above expression reduces to 
\be
\nabla^2 \delta \phi + \cF_4 \delta \phi  = 0 \,.
\ee
The functions  $\cF_\ast$ are given by
\be
\cF_\ast e^{-2 A} = \left( \frac{f_1}{16 f_4 -1} \,,   \frac{f_1}{16 f_4 -1}\,, \frac{f_2}{4 f_5 -1} \,, \frac{f_3}{16 f_4 -1} \right) \,.
\ee
Notice that the functions $\cF_\ast$ parametrises the mass of the variations. In order to study their stability we must show that all functions $\cF_\ast$ are positive, at least for some region on the parameter space of $\mathbf{S}^2$. \\

\subsubsection{An example}
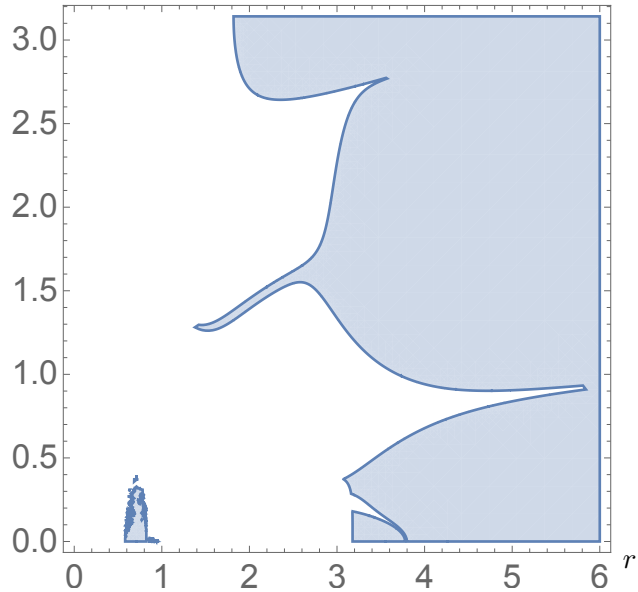
\begin{figure}[!h]
 \centering
 \includegraphics[width=8cm]{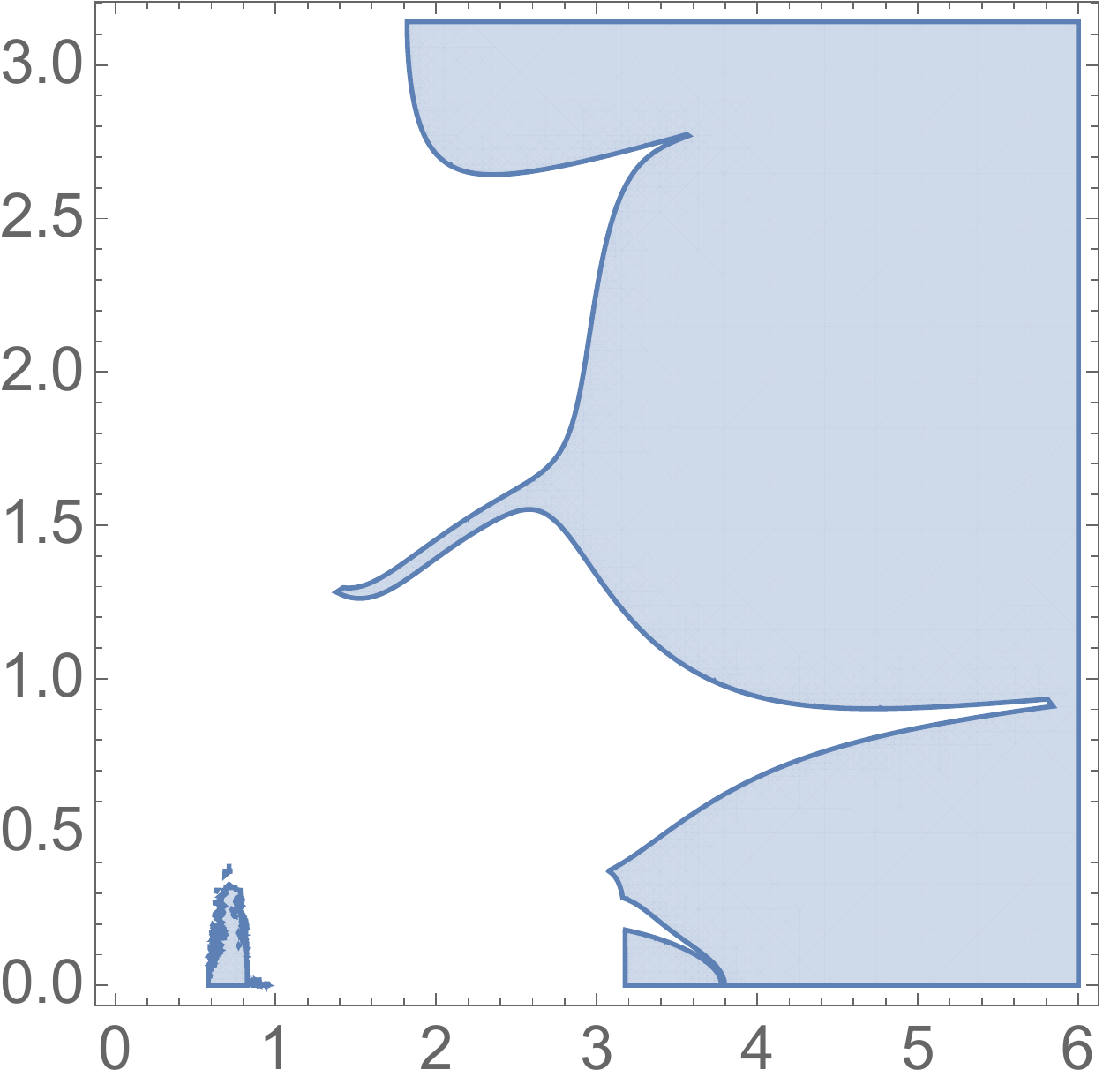}  
 \begin{picture}(0,0)
  \put(-237,230){$\theta$}
   \put(0,12){$r$}
  \end{picture}
\label{fig:st1}
\caption{Mass for variations $\delta A$, $\delta g$ and $\delta \phi$. Blue regions indicate that all mass are positive, while white regions are related to tachyonic variations.}
\end{figure}

Since a generic numerical analysis is beyond the scope of this work, we select a very simple case in which the holomorphic functions are specifically given by
\be
U_1 (z) = \log z \,, \quad U_2 (z) =  \log z + z \, \quad h(z) = z^{n} \,,
\ee
standing for the presence of branes at the singular point $z=0$, around which fluxes $B_2$ and $C_2$ get a monodromy. Notice that this example approximates the holomorphic functions to leading order on $z_p=0$. With the purpose to find stability, we explore the region $(z,\bar z)$ using the parametrization, $z = r e^{\rm{i} \theta}$ around the origin shown in Figure \ref{fig:st1}, where we find that indeed, there exists a region for which all variations posses a positive squared mass.

Although this particular example is far from realistic, we can see that for some internal parameters in $\mathbf{S}^2$, the $\phi$-perturbations are stable.

%-----------------------------------------------------------------------------------------------------------------------------------------------------
\section{Conclusions and Final Remarks}

By considering a specific non-trivial flux configuration composed by only 3-form fluxes and a real dilaton, we have studied conditions to solve four-dimensional Einstein's equations related to type IIB string compactification on a six-dimensional space given by the fibered product $T^4\times _z\mathbf{S}^2$,  were the two-dimensional sphere is punctured by the presence of 5- and 7-branes (charged under RR or NS-NS fields). We have found that such solutions exist by 
demanding the presence of  `meromorphic fluxes"  on $\mathbf{S}^2$.  By ``meromorphic fluxes" we refer to meromorphic functions which real part is given by non-trivial closed string potentials (NS-NS or RR) and imaginary component given by also non-trivial warping factors of the given background. By this simple request on the fluxes, we are able to prove also that Bianchi identity and tadpole conditions are fulfilled. \\

Moreover we observe that by considering meromorphic fluxes we can circumvent some standard results from a constant flux compactification. For instance, it is well known that in order to obtain a Minkowski vacuum it is required either the presence of only 1-form fluxes or $-$if one is considering 3-form fluxes$-$ it is necessary the presence of an orientifold 3-plane. Also, in such a case, it is understood that in the absence of 5-form fluxes, the warping factor must be  constant. In a meromorphic flux compactification (at least on $T^4\times _z\mathbf{S}^2$) it is possible to obtain a Minkowski vacuum with only 3-form fluxes and no orientifold 3-planes.\\

Contributions of fluxes and local sources to the trace of the energy-momentum tensor in  4D Einstein's equations vanishes due to the Global Residue Theorem, which states that the sum of residues on a compact space, $\mathbf{S}^2$ in our case, vanishes for meromorphic functions.\\

Based on these facts, we constructed a family of solutions of Einstein's equations and found that they correspond to a family of solutions constructed by the use of U-dualities  in G-theory and by fulfilling the Killing equations of motion preserving half of the supersymmetries in the effective model \cite{Candelas:2014jma,Candelas:2014kma}. Moreover, we observed that different solutions connected by U-dualities correspond to different solutions of Einstein's equations.\\

We then constructed supersymmetric solutions with different flux configurations for which we only required two different meromorphic functions $U$. Also we have computed the corresponding soft terms. It is important to notice that their behaviour is similar as those computed in a large volume compactification. We also presented a non-supersymmetric solution related to a compactification threaded by non-trivial NS-NS and RR 3-form fluxes. We showed that, by assuming spherical symmetry, this configuration can be stable for some regions of the moduli space of $\mathbf{S}^2$.\\

Notice that in order to construct the punctured two-dimensional sphere, we have made use of the results shown in \cite{Greene:1989ya}, where it is computed how branes curl up the complex plane by a deficit angle. In our case, they are 5- and 7-branes (some of them sourcing the 3-form fluxes and the real dilaton), and it is possible that some of them have negative charge and/or negative tension. Hence, although the presence of five or seven-dimensional orientifold planes are not discarded, orientifold 3-planes are. It will be desirable to consider more generic compactifications with fluxes depending on all 3 complex coordinates and look for generic conditions on which it is possible to obtain de Sitter and Minkowski vacua. \\

String compactification is the key scenario to connect string theory with four-dimensional physics. Although Calabi-Yau compactifications are exact vacuum solutions it was shown that more generic scenarios were required to step close to realistic effective models. Compactification with trivial fluxes has established a rich setup to explore the vast possibilities given by string theory. However it may be time to go further. We believe that compactification threaded by non-trivial fluxes, in particularly by meromorphic functions, is the next step towards more realistic models.

\vspace{1.0cm}

\begin{center}
{\bf Acknowledgements}
\end{center}
We thank Nana Cabo Bizet, Gustavo Niz and Miguel Sabido for useful explanations and nice discussions. We specially thank Ralph Blumenhagen and Francisco Morales for very important suggestions.  C. D. is supported by PRODEP with project number UGTO-PTC-524. O.L.-B. was partially supported by CONACyT under project No. 258982 and DAIP-CIIC 2016-2017.

\appendix
\section{Notation}
We follow the following index notation:
\begin{eqnarray}
\{M,N,P,...\}&:& \text{ label 10-dimensional space-time coordinates, from $0,\cdots , 9$}. \nonumber\\
\{\mu,\nu,...\}&:&\text{label 4-dimensional space-time coordinates from $0,\cdots , 3$}. \nonumber\\
\{m, n, ... \}&:&\text{label 6 dimensional coordinates of the compact space $T^4\times S^2$ from $1,\cdots , 6$}. \nonumber\\
\{i,j,k,...\}&:&\text{label coordinates of $T^4$ from $1, \cdots ,4$}.\nonumber\\
z_p&:& \text{singular points on $S^2$}\nonumber\\
\{a,b,...\}&:&\text{label spinorial indices in the compact space $T^4\times S^2$}.\nonumber\\
\{\alpha,\beta, \gamma, ...\}&:&\text{indices denoting the number of scalars in the soft term lagrangians}.\nonumber\\
\{\theta,...\}&:&\text{label the number of spinors in the soft lagrangian terms}.\nonumber\\
\{u,v,w,...\}&:&\text{complex coordinates on $T^4$.}\nonumber\\
\{I,J,K,...\}&:&\text{label D-brane's worldvolume coordinates}.
\end{eqnarray}

\section{Useful gamma identities}
The gamma matrices are defined in curved coordinates as
\be
\{ \gamma^m , \gamma^n \} = 2 g^{mn}
\ee
Some useful gamma identities useful are
\bea
\gamma^{m n} = \frac{1}{2} [ \gamma^m , \gamma^n ] \, , \nn \\
\gamma^{m n o} = \frac{1}{2} \{ \gamma^m , \gamma^{n o} \} \, , \nn \\
\gamma^{m n o p} = \frac{1}{2} [ \gamma^m , \gamma^{n o p} ] \, ,
\eea
we frequently use the eigenvalues of the gamma matrices with two indices acting on a chiral spinor. These
eigenvalues depends on the metric as
\be \label{eq:gamma_eig}
\gamma^{mn} \eta = (g^{mn}-g^{mm} g^{nn})^{1/2}\eta \, ,
\ee
And the factorization of gamma matrices
\bea
\gamma^{mnp} &= \gamma^{np} \gamma^{m} \quad m\neq n \neq p \, , \nn \\
\gamma^{mnpq} &= -\gamma^{npq} \gamma^{m} \quad m\neq n \neq p \neq q \, , \nn \\
\gamma^{mnp}\gamma_q &= \gamma^{m n} \delta^p_q +\gamma^{p n} \delta^n_q + \gamma^{n p}\delta^m_q -\gamma^{mnp}_q \, ,
\eea
where we can lower and upper indices by contraction with $g_{mn}$.

\section{Non-zero components of spin connection}
For the covariant derivative we employ the definition
\be
\nabla_m = \del_m + \frac{1}{4}\omega_{m \underline{a}\underline{b}}\gamma^{\underline{a}\underline{b}} \, ,
\ee
where as usual
\be
\omega_{\mu}^{\underline a \underline b} = 2 e^{\nu [\underline a}\del_{[\mu}e^{\underline b]}_{\nu]}-e^{\nu [\underline a}e^{\underline b]\sigma}e_{\mu \underline c}\del_{\nu}e^{\underline c}_{\sigma}
\ee
where the underliying variables are related with flat indices. As usual the metric and the components of
the vielbein are related by $g^{mn}=\eta^{\underline{mn}}e^{m}_{\underline{m}} e^n_{\underline{n}}$.\\

The non-zero spin-connection for the anzats  given in (\ref{eq:manzats}) are
\bea \label{eq:spin_1}
\omega_{\mu \underline{\mu z}} &= \frac{e^{A-D} \bar \del A}{|h|} \, , \quad \quad
\omega_{\mu \underline{\mu \bar z}} = \frac{e^{A-D} \del A}{|h|} \nn \, ,\\
\omega_{i\underline{i z}} &= \frac{e^{-D} \bar \del g_{ii}}{2 g_{ii}^{1/2}|h|} \, , \quad \quad
\omega_{i\underline{i \bar z}} = \frac{e^{-D}  \del g_{ii}}{2 g_{ii}^{1/2}|h|} \nn \, ,\\
\omega_{z\underline{z \bar z}} &= \frac{\del_z h}{2 h} + \del_z D \, , \quad
\omega_{\bar z\underline{z \bar z}} = -\frac{\bar \del_z \bar h}{2 \bar h} - \bar \del_z D \, .
\eea

%------------------------------------------------------------------
\section{Global Residue Theorem}
\label{GRT}

For completeness we reproduce the prove of the Global Residue Theorem \cite{Griffiths:1979pa}. Consider a meromorphic function $\omega(z)$ defined on a compact space with singularities denoted $\mathbf{M}$.  The singularities can  in principle be essential unremovable  or  $m$-order poles. In any case, according to Riemann's theorem, we can write the  function $\omega(z)$ as
\be
\omega(z)=\frac{g(z)}{f(z)},
\ee
where $g(z)$ is a holomorphic function with no singularities, and $f(z)$ tends to zero as $z$ approaches the singularities. The meromorphic function defines then a 1-form $\omega_1=\omega(z)dz$, which is closed in the space
\be
M_\ast=M-\sum_p {\cal U}_p,
\ee
where ${\cal U}_p$ is a vicinity of the singularity $z_p$. It follows then
\be
\omega_1\in \text{H}^1(\mathbf{M}^2/\sum {\cal U}_p, \mathbf{Z}).
\ee
The Global Residue Theorem states that the total sum of residues of $\omega$ on $M$ vanishes. This is proven as follows:
\be
\sum_p Res_p \omega =\sum_p \int_{\partial{\cal U}_p} \omega_1=\int_{\partial{M_\ast}}\omega_1=\int_{M_\ast}d\omega_1=0,
\ee
where in the last term we have used Stoke's theorem.

\subsection{Bianchi identity}
The Global Residue Theorem, tells us that  holomorphic functions on compact spaces with singularities integrate to zero over a closed non-contractible curve, or in other words, that  all residues related with those singularities sum up to zero. In our case
this also implies that Bianchi identities for $H_3$ and $F_3$ are globally fulfilled and thereof, total internal 5-brane charge vanishes. 
This guarantees that Bianchi identities are fulfilled for $F_3$ and $H_3$  while cancelling tadpoles for D5 and NS5-branes \footnote{Cancelation of NS tadpoles is significative since their presence establishes an obstacle to a satisfactory picture of supersymmetry breaking \cite{Dudas:2004nd}.}. \\

Here we want to show  how fluxes $H_3$ and $F_3$ satisfy Bianchi identity in the compact space $\mathbf{S}^2$ punctured by the presence of 5-branes. Since in the present work we are considering only 3-form fluxes from the RR and NS-NS potential given in (\ref{eq:ByC}), we have
\be
F_3=\sum_{ij}2 \text{Re}(f_1)\wedge dy^i\wedge dy^j,
\ee
where the complex  1-form $f_{1}$ is given by\footnote{Notice we have omitted the indices $i,j$ in $f_1$ just for simplicity.}
\be
f_{1}= (\partial_Z c_{ij})dz.
\ee
In the complex plane, with no singularities, Bianchi identity for $F_3$ is satisfied for $c_{ij}(z,\bar{z})$ being harmonic conjugate, or equivalently, as the imaginary part of a holomorphic function $U(z)$. However, once the complex plane has been compactified by addition of singular sources (branes sitting at points in $\mathbf{S}^2$) Bianchi identity for $F_3$ is altered. Here, we want to to show that indeed, in a compact space, $\int dF_3=0$  in spite of the presence of brane singularities. Notice that this implies tadpole cancelation for 5-brane charged sources.\\

It follows from the Global Residue Theorem \cite{Griffiths:1979pa} that
\be
\oint_{\partial U_p} f_1=\frac{1}{2\pi \rm{i}} \sum_p \text{Res}\left(\partial_z c_{ij}(z_p)\right)=0.
\ee
since
\begin{eqnarray}
\int_{\Sigma_4} dF_3 &=& \int_{\mathbf{S^2_\ast}}(df_1+\bar{d f_1}) \int_{\Sigma_{ij}}dy^i\wedge dy^j,\nonumber\\
&=& \sum_p\left(\oint_{\partial {\cal U}_{p}}f_1 +\oint_{\partial {\cal U}_{p}}\bar{f_1}\right)\int_{\Sigma_{ij}}dy^i\wedge dy^j,\nonumber\\
&=&0,
\end{eqnarray}
where in the last line we have used Stoke's theorem with $\Sigma_{ij}$ being a 2-cycle in $T^4$. Hence, although the local contribution of $dF_3$ is different from zero nearby a singular 5-brane location, the global contribution of all sources vanish. Bianchi identity for $F_3$ integrated over $\mathbf{S}^2$ implies tadpole cancelation for D5-branes wrapping 2-cycle in $T^4$ and extended in the 4-dimensional space-time. A similar analysis holds for the NS-NS flux $H_3$ related to NS5-branes sitting a point in $\mathbf{S}^2$. Notice as well that having a non-constant dilaton in our set up, implies the presence of 7-branes wrapping a 4-cycle in $T^4$ and extending as well in the whole 4-dimensional space-time. Both sets of branes, five- and seven-branes, can induce a D3-brane charge and interacting terms in the effective 4-dimensional action. \\

Now, in order to prove that integration on the internal space of $\hat{T}^l$ vanishes, we want to fix the values of the residues on each singularity compatible with integer flux quantization. Since the fluxes $H_{zij}=\partial_zb_{ij}$ are meromorphic functions on $S^2$ we can write by Riemann theorem that
\be
\partial_zb_{ij}=\sum_n \alpha_{n, ij} (z-z_p)^n + \sum_m \beta_{m, ij} (z-z_p)^{-m},
\ee
such that
\be
\int_{\partial U_p }\partial_zb_{ij} dz= \frac{1}{2\pi \rm{i}} \beta_1(z_p),
\ee
since the charge attributed to $H_3$ comes from a $NS5$-brane, one can choose that
\be
\beta_1(z_p)=2\pi \rm{i} (\mu_5+\rm{i}\nu_5)_k, \qquad \mu_5 \in\mathbf{Z}/2,
\ee
and then $\int H_3 \in\mathbf{Z}$. It follows from Bianchi identity of $H_3$ that total $NS5$-brane charge vanishes, this is
\be
\sum_{p=1} \beta_{1,p}=0,
\ee
and similarly for RR D5-brane charge.\\

%-----------------------------------------------------------------
\section{Effective DBI theory}
\label{DBI5-3}

Dirac-Born-Infeld theory for a set of $D5$-branes \cite{Myers:1999ps} is
\be
S^{D5}_{DBI}=\kappa_5\int d^5\sigma ~ e^{-\phi}\sqrt{det(\ast{\cal F}_{IJ})det Q^i_j},
\ee
where
\be
\ast{\cal F}_{IJ} = {\cal F}_{MN}D_IX^M \,D_J X^N
\ee
with $I,J$ running over the brane's worldvolume coordinates and ${\cal F}_{MN}=G_{MN}+B_{MN}$ and $D_I X^M = \partial_I X^M+ [ A_I , X^M ]$, where we are ignoring the presence of magnetic two-form fields on the $D5$-brane's world volume. It follows that
\be
(\ast {\cal F})_{IJ}= {\cal F}_{IJ}+{\cal F}_{Im}D_JX^M +{\cal F}_{mn}D_IX^m\partial_J X^n.
\ee
By taking a D5-brane wrapping a two-cycle in $T^4$, we can decompose the scalar fields $X^m$ in a Fourier series and keep the zero modes for which

\be
(\ast{\cal F})_{IJ}= \begin{bmatrix}
{\cal F}_{\mu\nu}+ {\cal F}_{mn}\partial_\mu X^m\partial_\nu X^n& &&b_{im}\partial_\nu X^m\\
0&&& {\cal F}_{ij}
\end{bmatrix}
\ee
At first order
\be
det(Q_J^I) = 1 - \frac{\sigma^2}{4}[X^I , X^J]^2 + ...
\ee
Therefore, the effective theory can be expressed as
\be
S^{D5}_{DBI}=\frac{\kappa_5}{\kappa_3}S^{D3}_{DBI_0} \left( 1 - \frac{\sigma^2}{4}[X^I , X^J]^2  + ... \right) \int d^2\xi \sqrt{det ({\cal F}_{ij})} .
\label{eq:DBI5-3}
\ee
where $DBI_0$ correspond to the abelian D3-brane action. Thus, at first order, in our anzats the DBI action decomposes into a D3 non-abelian part times a common factor.  According to our ansatz, the internal integral depends only on the complex coordinates ($z, \bar{z})$ of $\mathbf{S}^2_\ast$. It follows that soft masses can be computed using the DBI action for D3-branes.

\bibliographystyle{utphys}
\bibliography{references.bib}

\end{document}

%% file: mymacros.tex
\usepackage{amssymb, amsmath, amsthm,latexsym}
\usepackage[english]{babel} 
\usepackage[latin1]{inputenc} 
\usepackage{graphicx}
\usepackage[all]{xy}
\usepackage{bbm}
\usepackage{psfrag}

\newcommand{\be}{\begin{equation}}
\newcommand{\ee}{\end{equation}}
\def\bea#1\eea{\begin{align}#1\end{align}}
\newcommand{\nn}{\nonumber}

\newcommand{\w}{\wedge}

\def\del {\partial}

\newcommand{\cW}{\mathcal W}
\newcommand{\cV}{\mathcal V}
\newcommand{\cA}{\mathcal A}
\newcommand{\cB}{\mathcal B}
\newcommand{\cC}{\mathcal C}

\newcommand{\cF}{\mathcal F}

\newcommand{\cN}{\mathcal N}

\newcommand{\cM}{{\cal M}}